\documentclass[twocolumn,superscriptaddress,showpacs,amsfonts,amsmath,amssymb,prb]{revtex4}
\usepackage{graphicx} 
\usepackage{color}
\usepackage{epstopdf} 
\usepackage{url} 
\usepackage{multirow}
\usepackage{subfig, textcomp} 
\usepackage{dcolumn} 
\usepackage{bm} 
\graphicspath{{./FIGURES_4_arxiv/}}

\newcolumntype{M}{>{$\vcenter\bgroup\hbox\bgroup}c<{\egroup\egroup$}}

\begin{document}

\title{Absence of hyperfine effects in $^{13}$C-graphene spin valve devices}

\author{M. Wojtaszek} \affiliation{Physics of Nanodevices, Zernike Institute for Advanced Materials, University of Groningen, Groningen, The Netherlands} \email{m.wojtaszek@rug.nl} 
\author{I. J. Vera-Marun} \affiliation{Physics of Nanodevices, Zernike Institute for Advanced Materials, University of Groningen, Groningen, The Netherlands} 
\author{E. Whiteway} \affiliation{Department of Physics, McGill University, Montreal, Canada}
\author{M. Hilke} \affiliation{Department of Physics, McGill University, Montreal, Canada}
\author{B. J. van Wees} \affiliation{Physics of Nanodevices, Zernike Institute for Advanced Materials, University of Groningen, Groningen, The Netherlands}

\date{January the 14$^{\textrm{th}}$, 2013}

\begin{abstract} 
The carbon isotope $^{13}$C, in contrast to $^{12}$C, possesses a nuclear magnetic moment and can induce electron spin dephasing in graphene. This effect is usually neglected due to the low abundance of $^{13}$C in natural carbon allotropes ($\sim$1 \%). Chemical vapor deposition (CVD) allows for artificial synthesis of graphene solely from a $^{13}$C precursor, potentially amplifying the influence of the nuclear magnetic moments. In this work we study the effect of hyperfine interactions in pure $^{13}$C-graphene on its spin transport properties. Using Hanle precession measurements we determine the spin relaxation time and observe a weak increase of $\tau_{s}$ with doping and a weak change of $\tau_{s}$ with temperature, as in natural graphene.  For comparison we study spin transport in pure $^{12}$C-graphene, also synthesized by CVD, and observe similar spin relaxation properties. As the signatures of hyperfine effects can be better resolved in oblique spin-valve and Hanle configurations, we use finite-element modeling to emulate oblique signals in the presence of a hyperfine magnetic field for typical graphene properties. Unlike in the case of GaAs, hyperfine interactions with $^{13}$C nuclei influence electron spin transport only very weakly, even for a fully polarized nuclear system. Also, in the measurements of the oblique spin-valve and Hanle effects no hyperfine features could be resolved. This work experimentally confirms the weak character of hyperfine interactions and the negligible role of $^{13}$C atoms in the spin dephasing processes in graphene. 

\end{abstract}

%% http://www.aip.org/pacs/pacs2010/individuals/pacs2010_regular_edition/index.html %% 85.75.-d Magnetoelectronics; spintronics: devices exploiting spin polarized transport or integrated magnetic fields %% 85.75.Hh Spin polarized field effect transistors %% 72.80.Vp Electronic transport in graphene %% 72.25.-b Spin polarized transport (for spin polarized transport devices, see 85.75.-d) %% 72.25.Rb Spin relaxation and scattering %% 72.25.Dc Spin polarized transport in semiconductors
% 74.62.Dh	Effects of crystal defects, doping and substitution (for specific crystal defects, see 61.72.-y)
% 72.10.Fk	Scattering by point defects, dislocations, surfaces, and other imperfections (including Kondo effect)
% 68.43.-h	Chemisorption/physisorption: adsorbates on surfaces

\pacs{72.25.-b, 85.75.-d, 31.30.Gs} 
\maketitle 
%%%%%%%%%%%%%%%%%%%%%%%%%%%%%%%%%%%%%%%%%%%%%%%%%%%%%%%%%%%%%%%%%%%%%%%%%%%%%%%%%%%%%%%%%%%%%%%%%%%%%%% %%%%%%%%%%%%%%%%%%%%%%%%%%%%%%%%%%%%%%%%%%%%%%%%%%%%%% 

\section{\label{sec:Introduction}Introduction} 

Spin transport in graphene has attracted a lot of research attention due to predictions of high spin relaxation times $\tau_{s}$ and large spin relaxation lengths $\lambda_{s}$ \cite{Huertas-Hernando2006}. The experimentally determined values of $\tau_{s}$ (Refs.~\onlinecite{Tombros2007, Han2011,  Guimaraes2012}), center around values of 100 ps - 1 ns, three orders of magnitude lower than expected. This discrepancy between theory and experiment motivates the need to identify the mechanisms for spin dephasing \cite{Zhang2012, Kochan2014}. 

A well-known source of dephasing is the presence of random magnetic moments (for example from localized states \cite{McCreary2012, Birkner2013}). The $^{13}$C isotope with nuclear spin $I_{N}=\frac{1}{2}$ also possesses a magnetic moment but this is usually neglected due to the low abundance of $^{13}$C in natural carbon allotropes ($\sim$1 \%) and a weak hyperfine coupling of $\lesssim$0.6 $\mu$eV, about 100 times smaller than for GaAs \cite{Fischer2009, Paget1977}. Although there are many theoretical evaluations of the size of hyperfine interactions \cite{Fischer2009, Yazyev2008, Dora2010} and their role in spin transport, they lack experimental verification in graphene.

% an aspect not explored so far in graphene spin transport. 
%The strength of the interactions between the electron spins and nuclear spins, so called hyperfine coupling strength, is predicted to be very small  in graphene and in carbon nanotubes ($\lesssim$0.6 $\mu$eV), about 100 times smaller than for GaAs \cite{Fisher2009, Paget1977}. However, experimental studies of pure $^{13}C$-based carbon nanotubes \cite{Churchill2009} report large enhancement of hyperfine interactions, in discrepancy with the theoretical work. 
% As the spin accumulations in graphene can be quite high then even a weak hyperfine coupling may overall effectively affect the spin of conducting electrons.

In this work we demonstrate spin transport in pure $^{13}$C-isotope graphene and compare it with spin transport in pure $^{12}$C-isotope graphene using the non-local spin-valve geometry. We use Hanle precession measurements to characterize the spin properties at room temperature and at 4.2~K for different carrier densities. We also amplify the hyperfine effects by increasing the spin polarization in graphene, to induce dynamical nuclear polarization (DNP). The depolarizing effect of nuclei is best observed under an oblique external magnetic field, which makes the orientation of nuclear spin non-collinear to the electron spin, causing extra spin precession. To quantify this effect on the spin signals we model the non-local spin-valve and Hanle precession effects at oblique angles for various degrees of graphene polarization. The estimated hyperfine features in spin transport are below the experimentally achievable resolution, which we confirm later experimentally. These measurements are reproduced in several graphene regions as well as in independently fabricated samples. By exploring the extreme conditions of pure $^{13}$C composition and high graphene polarization, we experimentally verify the weak character of hyperfine interactions in graphene and the negligible role of $^{13}$C atoms in spin dephasing in graphene. 

\section{\label{sec:Sample_Fabrication}Sample Fabrication} % 
The advent of synthetic methods to grow graphene \cite{Lin2012, Li2009, Wang2010} allows for chemical growth of graphene with an arbitrary composition of carbon isotopes. A pure $^{13}$C-graphene monolayer is synthesized on commercial Cu foil using chemical vapor deposition (CVD) from 99.9\% pure $^{13}$C-methane (CLM-3590-1, from Cambridge Isotopes Laboratories, Inc.) as described in Ref.~\onlinecite{Bernard2012}. 
Next, to transfer the graphene to an insulating substrate, we attach the graphene on Cu foil to a  polydimethylsiloxane stamp (PDMS) and dip it into FeCl$_{3}$ aqueous solution (1~g ml$^{-1}$) to etch away the copper. After the removal of Cu and subsequent dipping in deionized water to clean off the etching residues,  we transfer the graphene onto a 500-nm-thick layer of SiO$_{2}$ with a highly-doped Si substrate below to serve as a back gate. 

A homogeneous, single-layer graphene area is selected based on optical contrast and Raman spectroscopy \cite{Ferrari2006}, using a 532~nm laser. The Raman spectrum of pure $^{13}$C-graphene is shown in Fig.~\ref{fig:Raman_C12vsC13}, where for comparison we also show the spectrum of pure $^{12}$C-graphene. The latter is also a good representative of the spectrum of natural graphene, which has only $\sim 1\%$ of $^{13}$C abundance. When compared to $^{12}$C-graphene, the vibrational Raman modes in $^{13}$C-graphene display a downward shift  \cite{Bernard2012}: here from $\sim$1585 to 1525~cm$^{-1}$ for the $G$ band and from $\sim$2680 to 2580~cm$^{-1}$ for the $2D$ band. This shift arises from the difference in the atomic masses of the carbon isotopes and is consistent with the classical model of a harmonic oscillator, where its vibrational modes are inversely proportional to the square root of its masses. Raman spectroscopy also confirms the good quality of the selected graphene area, by the absence of a $D$ band in the spectrum~\cite{Pimenta2007}. 

After selecting a graphene region we define a rectangular strip of graphene using electron beam lithography and O$_{2}$ plasma etching. Then we define contacts using an $e$-beam in poly(methylmethacrylate) (PMMA) resist. First we evaporate 0.8~nm of Al and then naturally oxidize it, in order to turn it into a tunneling barrier. Next, we evaporate Co (30~nm) and 2~nm of Al on top as a capping layer to protect the cobalt from oxidation. The tunneling barrier of AlO$_{x}$ is present only underneath the contacts. All samples are measured in high vacuum. Low-temperature measurements are performed in a flow cryostat with a rotatable magnet around the in-plane and out-of-plane axes of the sample.

\section{\label{sec:Spin_Charge_Transport}Charge and spin transport in $^{13}$C-graphene.} % s
\begin{figure}
\centering
  \subfloat[]{  
   \includegraphics[width=0.5\columnwidth]{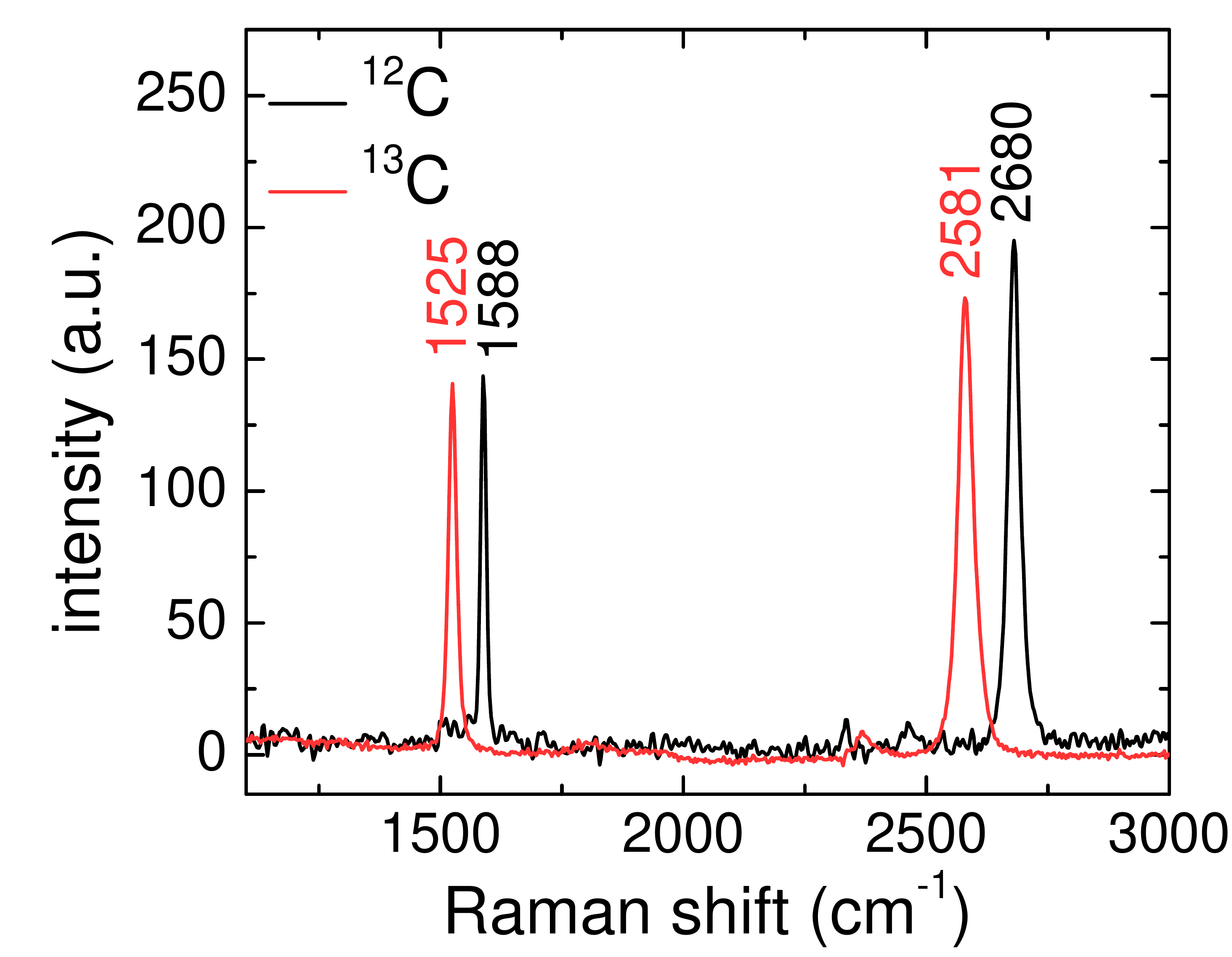} 
   \label{fig:Raman_C12vsC13}}
  \subfloat[]{  
   \includegraphics[width=0.5\columnwidth]{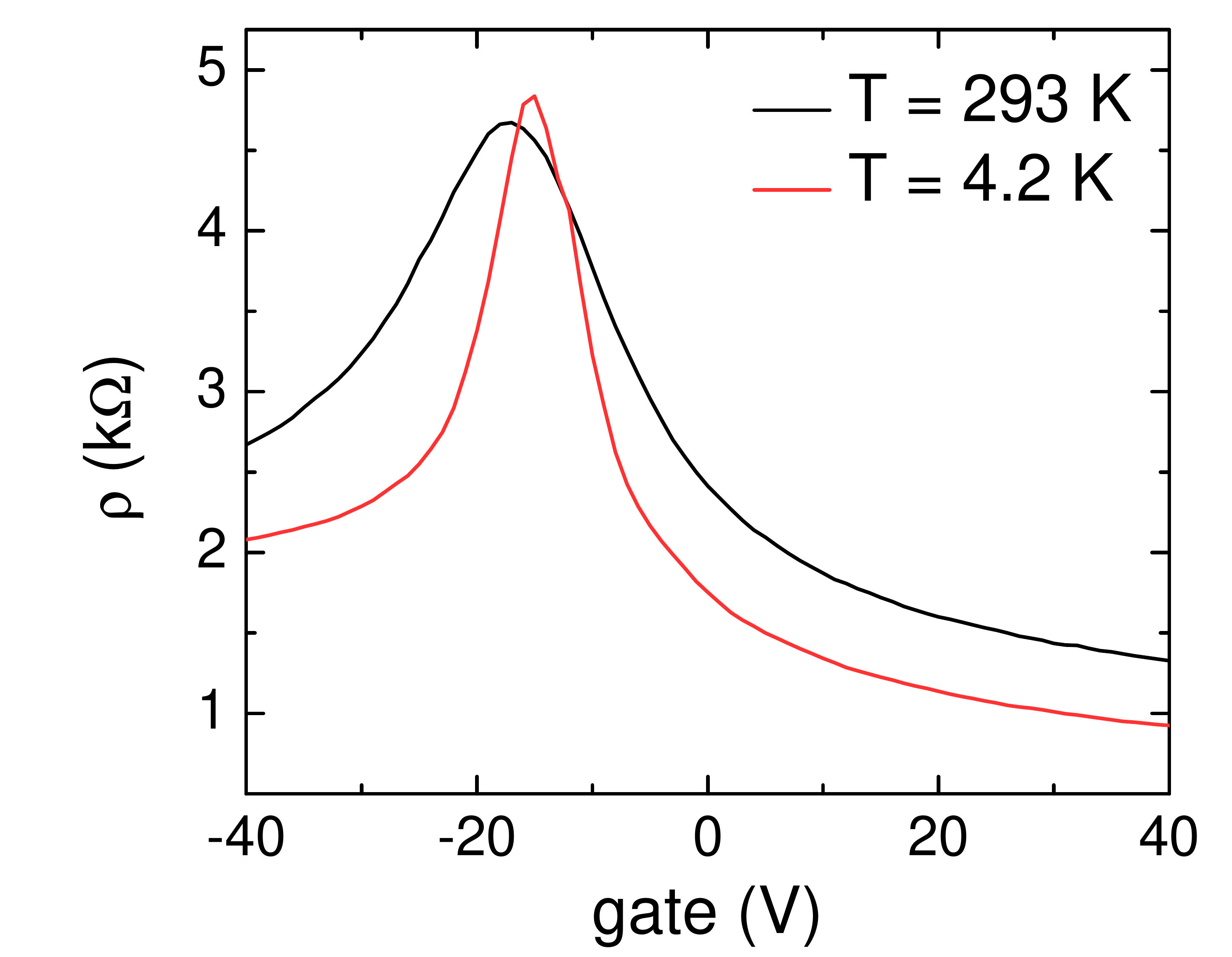} 
   \label{fig:Dirac_RTvsLT}}
\captionsetup{format=hang,justification=centerlast, font=small}
\caption[]{(Color online) (a) Raman spectrum of CVD graphene on SiO$_{2}$ of pure $^{12}$C (black) and pure $^{13}$C (red) isotopic content. The ratios of the Raman shift $\nu$ for graphene G and 2D bands reflect the difference in the mass of these isotopes: $\nu_{^{12}C} / \nu_{^{13}C} = \sqrt{13/12}$. (b) The typical curve of $^{13}$C-graphene sheet resistivity as a function of gate voltage at room temperature (black) and at T~=~4.2~K, (red). 
}
\end{figure}

Initially we characterize the sheet resistance $\rho$ of graphene as a function of the gate bias $V_{g}$ in a four-terminal measurement. The induced carrier concentration $n$ is calculated from $n = C_{g}(V_{g}-V_{0})/e$, where $V_{0}$ is the voltage corresponding to the maximum of $\rho$ (Dirac point), and $C_{g}$ is the gate capacitance, $C_{g} =$~70~aF/$\mu \text{m}^{2}$ for 500~nm SiO$_{2}$. The measured samples display similar electronic quality to micromechanically cleaved graphene \cite{Tombros2007, Jozsa2009}, with mobilities $\mu=(en \rho)^{-1}$ between 1000 and 3000 cm$^{2}$/Vs. As in exfoliated graphene on SiO$_{2}$ the maximum of $\rho(V_{g})$ does not vary strongly with temperature and only the Dirac peak displays a narrowing of its width due to the reduced thermal broadening; see Fig~\ref{fig:Dirac_RTvsLT}. 

\captionsetup[subfloat]{captionskip=-1.7em,margin = 0.1em,justification=raggedright,singlelinecheck=false,font=normalsize, position=top}

\begin{figure}

 \subfloat[]{\qquad  \includegraphics[width=0.435\columnwidth]{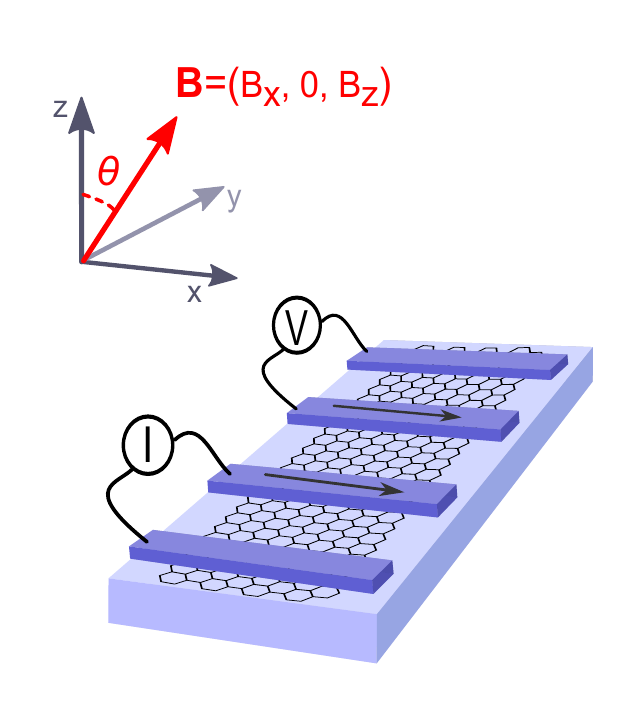} 
        		 \label{fig:NLSV_scheme}}
 \subfloat[]{  \includegraphics[width=0.5\columnwidth]{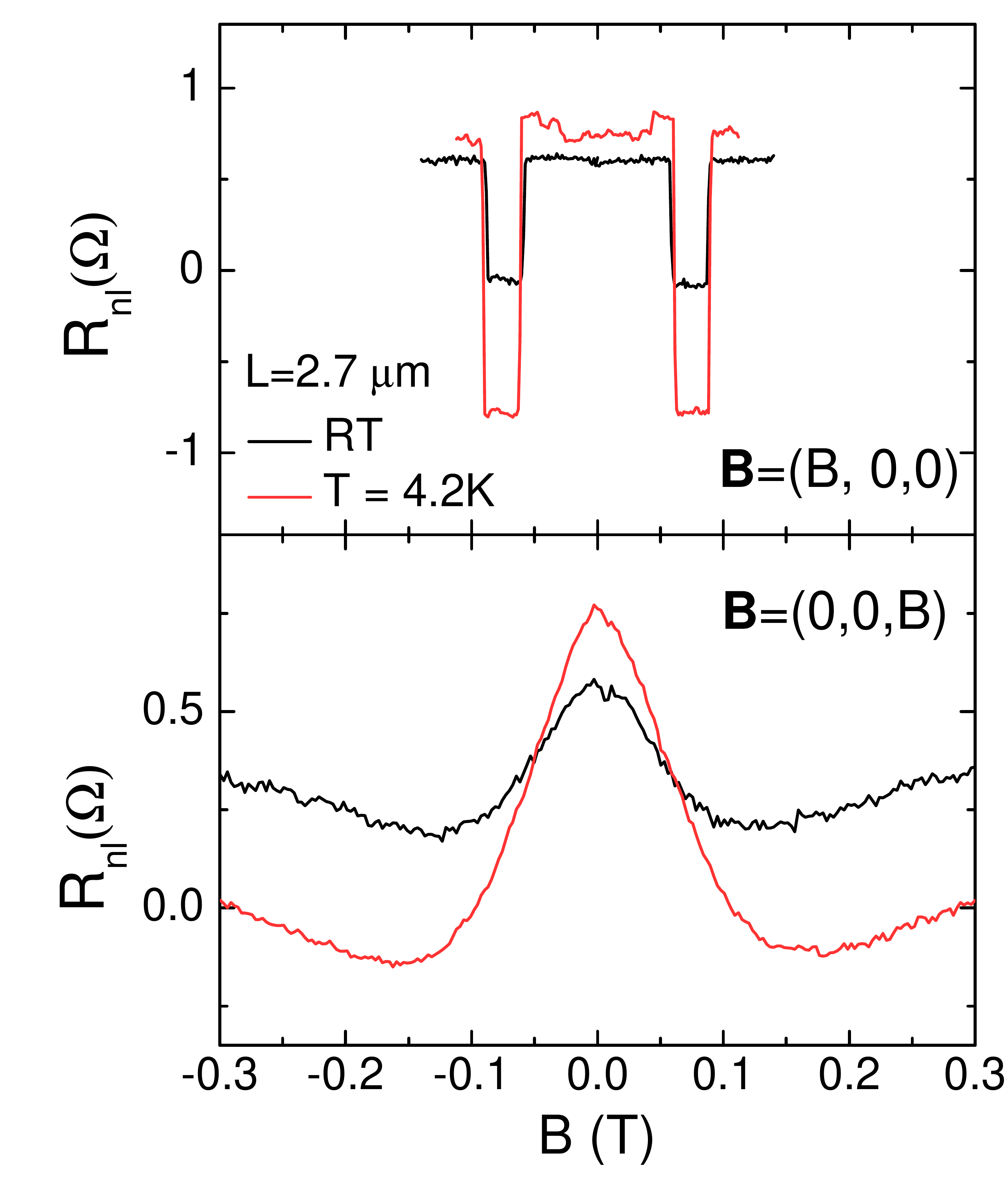} 
          		\label{fig:SVHanle_RT_LT}}
\captionsetup{format=hang,justification=centerlast, font=small}
\caption[]{(Color online) (a) Nonlocal detection scheme in a graphene spin-valve device. The arrows on the inner contacts mark their magnetization (here parallel). We measure signals for three different configurations of external field: in-plane $\mathbf{B}=(B, 0, 0)$ for the spin-valve effect, normal to the plane $\mathbf{B}=(0, 0, B)$ for regular Hanle effect, and at an angle $\theta$: $\mathbf{B}=(B\sin \theta, 0, B\cos \theta)$ for oblique Hanle effect. 
(b) Spin-valve and Hanle measurements in $^{13}$C-graphene at room temperature (black) and at T~=~4.2~K (red). The distance between ferromagnetic injector and detector is $L=$~2.7~$\mu$m and the contact magnetization is parallel. }
\end{figure}

Next we perform spin transport measurements in a non-local spin-valve geometry. In such a measurement we inject a spin polarized current through a ferromagnetic contact, the injector, and probe it with another ferromagnetic contact, the detector. The injection and detection circuits are separated [see Fig.~\ref{fig:NLSV_scheme}] which reduces the magnetoresistive background and electrical noise.

%%
%%\begin{tabular}{c c}
%%           \multirow{2}{*}{
%%            	\subfloat[]{ \qquad
%%      			\includegraphics[width=0.39\columnwidth]{drawing_device_FINAL.pdf} 
%%        		 \label{fig:NLSV_scheme}}}
%%            & { 
%%           		\subfloat[]{  
%%         		\includegraphics[width=0.5\columnwidth]{PUBL_RT_SV_Hanle_C.pdf} 
%%          		\label{fig:SVHanle_RT}}} \\
%%            &{
%%             	\subfloat[]{
%%         		\includegraphics[width=0.5\columnwidth]{PUBL_LT_SV_Hanle_D.pdf} 
%%         		\label{fig:SVHanle_LT}}}
%%        \end{tabular}
%%        
        
\captionsetup[subfloat]{captionskip=-1.7em,margin = 0.1em,justification=raggedright,singlelinecheck=false,font=normalsize, position=top}

A non-local spin resistance, defined as $R_{\text{nl}} = V_{\text{nl}}/I$, displays a switching "spin-valve" behavior when in-plane magnetic field is swept, correlated with the switching of the relative magnetization of the injecting and detecting contacts from parallel ($\uparrow \uparrow$) to antiparallel ($\downarrow \uparrow$) alignment. By setting the magnetic field perpendicular to the graphene plane we can study the spin precession (Hanle effect) \cite{Tombros2007, Jozsa2009, Han2011}. Typical measurements of the spin-valve and Hanle precession signals are presented in Fig.~\ref{fig:SVHanle_RT_LT} for room and  for liquid-helium temperature. At low temperatures the amplitude of the signal increases, but its features remain the same. To remove the spin-independent background we record Hanle curves for the $\uparrow \uparrow$ and $\downarrow \uparrow$ cases, where the pure spin signal is $R_{\text{nl}}=(R_{\text{nl}}^{\uparrow \uparrow}-R_{\text{nl}}^{\downarrow \uparrow})/2$, and is further used for fitting the spin coefficients.   

The Hanle effect can be described by the one-dimensional Bloch equation for the spin chemical potential $\boldsymbol{\mu_{s}}$: 
\begin{equation} 
   D_S\nabla^{2} \boldsymbol{\mu_s} - \frac{\boldsymbol{\mu_s}}{\tau_s} + \frac{g\mu_{B}}{\hbar} \boldsymbol{\mathbf{B}}\times\boldsymbol{\mu_s}=\mathbf{0} 
\label{eq:Bloch} 
\end{equation} 
which includes spin diffusion: the term with $D_{s}$, spin relaxation, the term with $\tau_{s}$, and spin precession, the term with magnetic field $\mathbf{B}$ where $g=2$ is the gyromagnetic factor of a free-electron and $\mu_{B}$ is the electron Bohr magneton. 
By fitting the Hanle curve to the solution of Eq.~\ref{eq:Bloch} we can independently determine the spin diffusion length $D_{s}$ and spin relaxation time $\tau_{s}$. For a more accurate extraction of spin coefficients we always make sure that the distance between the injector and detector $L$ is larger than the spin relaxation length $\lambda_{s}=\sqrt{\tau_{s} D_{s}}$, as motivated in Ref.~\onlinecite{Maassen2012}.

A full characterization of the spin properties at room and liquid helium temperatures as a function of carrier concentration $n$ is given in the Appendix \ref{sec:Spin_valve_polarization_changes}. 
Typically the values for $\tau_{s}$ in $^{13}$C-graphene range from 60 to 100~ps, depending on the sample doping.
The $\tau_{s}$ achieved are roughly twice smaller than previously reported for CVD graphene\cite{Avsar2011} and exfoliated graphene \cite{Tombros2007, Jozsa2009}, although the electron mobility in our samples is comparable. A lower $\tau_{s}$ can originate from structural defects and rippling of the graphene sheet\cite{Avsar2011, Huertas-Hernando2006, Zhang2012a}, which are inherent to the growth conditions, the quality of the catalytic substrate (Cu foil), and the transfer methods, as well as from eventual contamination with FeCl$_{3}$ etchant. 
The lower values of $\tau_{s}$ are also found in the control sample - a CVD-grown graphene from pure $^{12}$C precursor (see the next section), and therefore cannot be attributed to the dephasing from  hyperfine fields from the $^{13}$C nuclei.

%%%%%CHAKED UP TO THIS POINT
\section{\label{sec:Spin_C12_vs_C13} Spin transport properties in $^{13}$C versus $^{12}$C graphene.}
At $B \leqslant 0.5$~T and $T \simeq 300$~K the nuclei are randomly oriented so that spin dephasing can happen due to these fluctuating, weak nuclear moments. The effect of randomly fluctuating nuclear moments can be evaluated by comparing the spin properties between pure $^{13}$C- and pure $^{12}$C-graphene in room-temperature spin transport. A pure $^{12}$C-graphene monolayer is synthesized on Cu by CVD from 99.99\% pure $^{12}$C-methane  and then we follow the same device fabrication steps as for the $^{13}$C samples. The magnetic moment of the $^{13}$C nuclei \cite{Harris1978} is $\mu_{^{13}\text{C}}$ = 0.7$\mu_{n}$, where $\mu_{n}=e\hbar/M$ is the nuclear magneton. As $\mu_n$ is about 1800 times smaller than $\mu_{B}$, due to the much larger proton mass $M$, we have $\mu_{^{13}\text{C}}\simeq \mu_{B}/2600$. 

\begin{figure}
\centering
   \includegraphics[width=0.9\columnwidth]{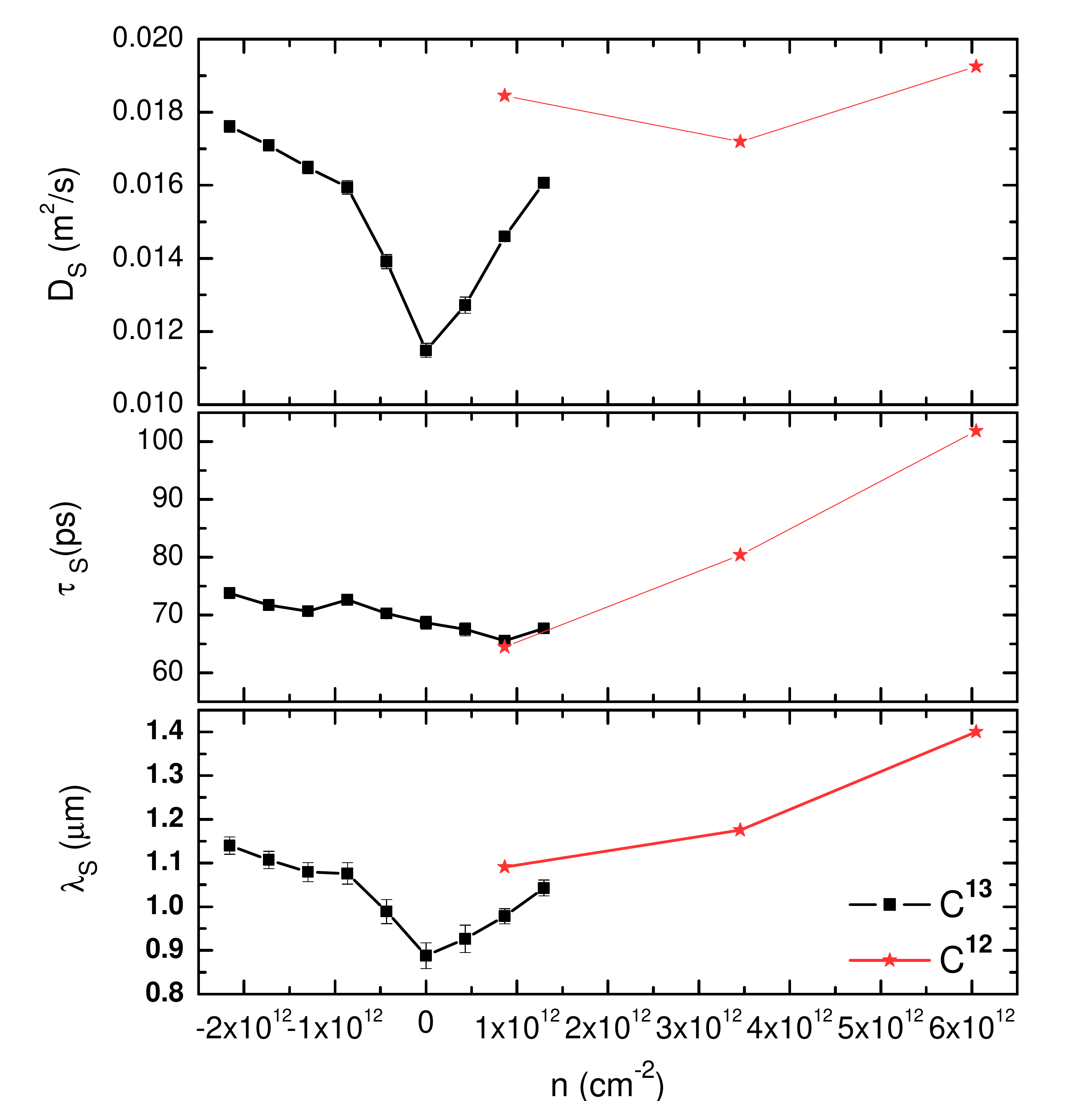} 
\captionsetup{format=hang,justification=centerlast, font=small}
\caption[]{(Color online) Comparison of spin properties at different carrier concentration $n$ between pure $^{13}$C and pure $^{12}$C graphene at room temperature. The coefficients are obtained from fitting Hanle measurements to the solution of Eq.~\ref{eq:Bloch}. For $^{13}$C we analyze data for injector-detector spacing $L$~=~2.7~$\mu$m and for $^{12}$C $L$~=~4~$\mu$m. }
 \label{fig:Comp_C13vsC12_RT}
\end{figure}

A comparison of room-temperature spin properties determined from Hanle fitting at different carrier concentration $n$ for these two isotopically pure graphenes is presented in Fig.~\ref{fig:Comp_C13vsC12_RT}. For $^{13}$C we analyze data for injector-detector spacing $L = 2.7 \mu$m and for $^{12}$C $L = 4 \mu$m, both longer than $\lambda_{s}$. The large $p$ doping in the $^{12}$C device enables us to record the spin properties only in the hole regime. The values for $D_{s}$, $\tau_{s}$, and $\lambda_{s}$ for both $^{13}$C and $^{12}$C are very similar, proving the negligible effect of random nuclei on electron spin transport. Here, we experimentally verify that the random, unpolarized nuclei do not contribute to the spin dephasing. In the next sections we will analyze the experimental situation when the nuclear polarization could be built up coherently. 

\section{\label{sec:oblique_COMSOL} Tracking coherent nuclear fields by an oblique Hanle effect.}
Although the $^{13}$C nuclei carry a smaller magnetic moment than the electrons, they outnumber them. In graphene the density of nuclei is $N=3.32\times 10^{15}$~cm$^{-2}$, so for pure $^{13}$C-graphene the product $N \mu_{I}$ is comparable to $n \mu_{B}$ for the density of conducting electrons, $n=1.3\times 10^{12}$~cm$^{-2}$, and greater for $n$ closer to the Dirac point. This means that once these nuclei are coherently polarized, they can produce a sizable nuclear magnetic field $\mathbf{B_{n}}$ and be a source of spin dephasing.  $\mathbf{B_{n}}$ adds vectorially to the external magnetic field $\mathbf{B}$ and can modify the line shape of the Hanle curve, for example by changing its width or the position of the maximum or inducing an asymmetry versus the external magnetic field \cite{Paget1977, Chan2009, Salis2009}.

%Although the theoretical values of hyperfine interactions in graphene are very small, of the order of 1~$\mu eV$~\cite{Fischer2009}, the experimental estimations in carbon nanotubes report hyperfine interactions of the order of 100~$\mu eV$~\cite{Churchill2009}, hence their effect might be measurable. 
First, we provide an estimate for nuclear effects and model the unique features of the electrical spin signal due to the presence of $\mathbf{B_{n}}$. We can neglect the Zeeman splitting as a source of nuclear polarization ($\mathbf{B}\leqslant 0.5$~T, $T\simeq 4.2$~K, so $|E_{Z}|=\gamma_{^{13}\text{C}} \hbar B < 25$~neV, where $\gamma_{^{13}\text{C}}=6.73 \times 10^{7}$~rad/Ts is the gyromagnetic ratio of the $^{13}$C isotope). Hence the nuclear polarization can arise only from the "flip-flop" angular momentum exchange with polarized electrons (DNP) \cite{Paget1977}. After such an angular momentum transfer the magnetic field produced by polarized nuclei (the Overhauser field) has the same orientation as the electronic spins so no hyperfine dephasing is induced. However, when one applies the external magnetic field at an oblique angle, the nuclear spin will immediately precess around the total magnetic field $\mathbf{B}+\mathbf{B_{e}}$, where $\mathbf{B_{e}}$ is the average magnetic field created by electrons (the Knight field). Following Paget~\textit{et al.}, \cite{Paget1977} we can express the nuclear field $\mathbf{B_{n}}$  as
\begin{equation}
\mathbf{B_{n}} = f b_{n} \frac{ (\mathbf{B}+ \mathbf{B_{e}}) \cdot \langle \mathbf{S}\rangle (\mathbf{B}+\mathbf{B_{e}}}{(\mathbf{B}+ \mathbf{B_{e}})^{2}}
\label{eq:Bnuclei}
\end{equation} 
where $\mathbf{B_{e}}=b_{e}\langle \mathbf{S}\rangle$, $\langle \mathbf{S}\rangle$ is the average  electron spin polarization  ($\vert\langle \mathbf{S} \rangle\vert=\frac{1}{2}$ for a fully polarized system), $b_{n}$ and $b_{e}$ are describing the effective magnetic fields produced by the nuclear and electron spin, respectively in the case of their complete polarization, and $f=T_{1}/(T_{1}+T_{1e})\lesssim 1$ is the leakage factor, which relates the spin relaxation due to hyperfine interactions  between the nuclei and the fluctuating magnetic field of the electrons $T_{1e}$ and other relaxation processes $T_{1}$ \cite{Chan2009, Dyakonov1984}. 
% nuclear polarization is a result of generation of non-equilibrium electron spin  by FM contact and depends on average electron spin S, which in turn is a function of the nuclear field BN. This os equivalent to the action of internal feedback in the electron- nuclear spin system. The polarization of the nuclear system is governed by the external magnetic field and electron polarization which is in turn a function of external magnetic field, nuclear polarization and polarization of injector. 
% if S=!0 nuclear spins are affected by an average electron field:
% Be= be*S, where be=-(16pi/3)*mu_0*ne zeta^2, ne=electron concentration, zeta=parameter characterizing the localizatio of the electron wave function on the nuclei.
% in turn polarized nuclei give rise to an effective nuclear field affecting spins:
% BN = bn Iav/I (I=spin of the nucleous => Iav/I is a polarization of nuclei like S is a polarization of electrons)
% bn=(16pi/3)*zeta^2*N*mu_n/g, N is the concentration of the nuclei
% ASSUMPTION: semiconductor of A3B5type, 
% here formulas for Bn, Be are when hyperfine interactions is the same for all lattice nuclei (true for CONDUCTION ELECTRONS). When electron is localized one needs to ionclude the distance from nucleus and localized center. Then be, bn depend on the distance (formulas of Paget). 
% in semiconductors as a rule N/ne>> mu_0/mu_n and therefore bn>>be.
%Besides the regular field Be the nuclei are affected by fluctuations of the electron field. These fluctuations transfer angular momentum of oriented electron to the nuclear spin system. 
From formula (\ref{eq:Bnuclei}) we see that $\mathbf{B_{n}}$ is proportional to the average electronic spin $|\langle \mathbf{S} \rangle| = \beta \times \frac{1}{2}$, where $\beta$ is the electron polarization or the ratio between the number of polarized carriers and the total number of carriers,  $\beta \simeq \frac{2\mu_{s}\cdot \nu(E_{F})}{n(E_{F})}$, where $\nu(E)$ is the density of states of graphene and $n$ is the number of states at a given Fermi level $E_{F}$. In the limit of zero-temperature: $E_{F}=\sqrt{\pi n} \hslash v_{F}$, where $v_{F}$ is the Fermi velocity of graphene,  and one gets $\frac{\nu(E_{F})}{n(E_{F}}=\frac{1}{E_{F}}$. As $\mu_{s}= e\Delta V_{\text{nl}}/P$ where $\Delta V_{\text{nl}}=(V_{\text{nl}}^{\uparrow \uparrow}-V_{\text{nl}}^{\downarrow \uparrow})/2$ and $P$ is the polarization of the detector, we can directly relate the non-local signal to graphene polarization: $\beta=\frac{2 e \Delta V_{\text{nl}}}{\sqrt{\pi n} \hslash v_{F} P}$.  

With both internal and external, non-collinear magnetic fields the Bloch equation [Eq.~(\ref{eq:Bloch})] now requires a full vectorial treatment. The mutual, non-linear dependence  of $\mathbf{B_{n}}$ and $\boldsymbol{\mu_{s}}$ requires solving the Bloch equation self-consistently. For that we choose a finite-element method package (COMSOL)\cite{Slachter2011}. 
We define the problem as time independent; therefore we assume that the experimental time scale is longer than the time necessary for the nuclei to adapt to the external magnetic field ($\sim$100~$\mu$s from the typical linewidth of the NMR $^{13}$C spectrum, \cite{DeBoer1974}). 
In Eq. (\ref{eq:Bnuclei}) we can distinguish two regimes: (1) for small external fields, $b_{e} \langle \mathbf{S} \rangle > \mathbf{B}$, when $\mathbf{B_{n}}$ is almost aligned with $\langle \mathbf{S} \rangle$, and (2) for large external fields, $b_{e} \langle \mathbf{S} \rangle \ll \mathbf{B}$, when $\mathbf{B_{n}}$ is almost aligned with $\mathbf{B}$. 
For the first regime a dephasing feature of nuclear origin can be observed in spin-valve measurements as a dip around $B = 0$ if we add to the sweeping in-plane field a small out-of-plane component $B_{z}$. In the second field regime  $\mathbf{B_{n}}$ can be identified using an oblique Hanle effect, where it leads to an asymmetry in precession curves and the appearance of additional satellite peaks. These features are best observed at large $L$ because the Hanle line shape can be fully recorded within a smaller field range, avoiding the switching of the magnetization of the contact. 

First we test our model for the case of doped GaAs, where the hyperfine effects are well understood~\cite{Paget1977, Chan2009, Salis2009}, and the effective coefficients $b_{n}$ and $b_{e}$ are known. As the model emulates properly all the features related to $\mathbf{B_{n}}$ in GaAs, see Appendix \ref{sec:Modeling_GaAs}, we can now apply it also to the case of $^{13}$C-graphene. However, there are some important remarks. In GaAs all the constituent isotopes: $^{69}$Ga, $^{71}$Ga and $^{75}$As have larger magnetic moments ($I_{N}=\frac{3}{2}$) and the hyperfine interactions are stronger (~90~$\mu$eV)\cite{Paget1977}, mainly due to the non-zero amplitude of the $s$ orbital of the electrons at the position of the nuclei (Fermi contact). In graphene, where the conducting electrons are of $\pi$ type, this amplitude is zero,\cite{Fischer2009} and only the much smaller anisotropic hyperfine term  $-0.3\mu$~eV (along the direction of electronic polarization) remains. This corresponds to $b_{n}\simeq -5.2$~mT, which is about 1000 times smaller than for GaAs \cite{Chan2009}. Additionally, for conducting electrons one can ignore the term $b_{e}$ \cite{Paget1977}. Therefore for graphene Eq.~(\ref{eq:Bnuclei}) simplifies to:
\begin{equation}
\mathbf{B_{n}} = f b_{n} \frac{ \mathbf{B} \cdot \langle \mathbf{S}\rangle \mathbf{B}}{\mathbf{B}^{2}}
\label{eq:Bnuclei_gr}
\end{equation} 
%(in InSb, where hyperfine interactions are strong, $b_{e}<$~0.1~mT for 10$^{15}$ polarized conducting electrons \cite{). 
Another important difference between graphene and GaAs is the time for building up the dynamical polarization. In doped GaAs the spin relaxation time $T_{1e}$ is faster for localized electrons  on donor ($T_{1e} = 0.1$~s) than for delocalized conduction electrons ($T_{1e} = 10^{4}$~s)\cite{Dyakonov1984}. DNP happens on the same timescale as $T_{1e}$ (because it is a reciprocal process to relaxation). As graphene primarily lacks localized states, we expect that one needs hours to build up DNP by solely conduction electrons.  

\begin{figure}
\centering
  \subfloat[]{  
   \includegraphics[width=0.5\columnwidth]{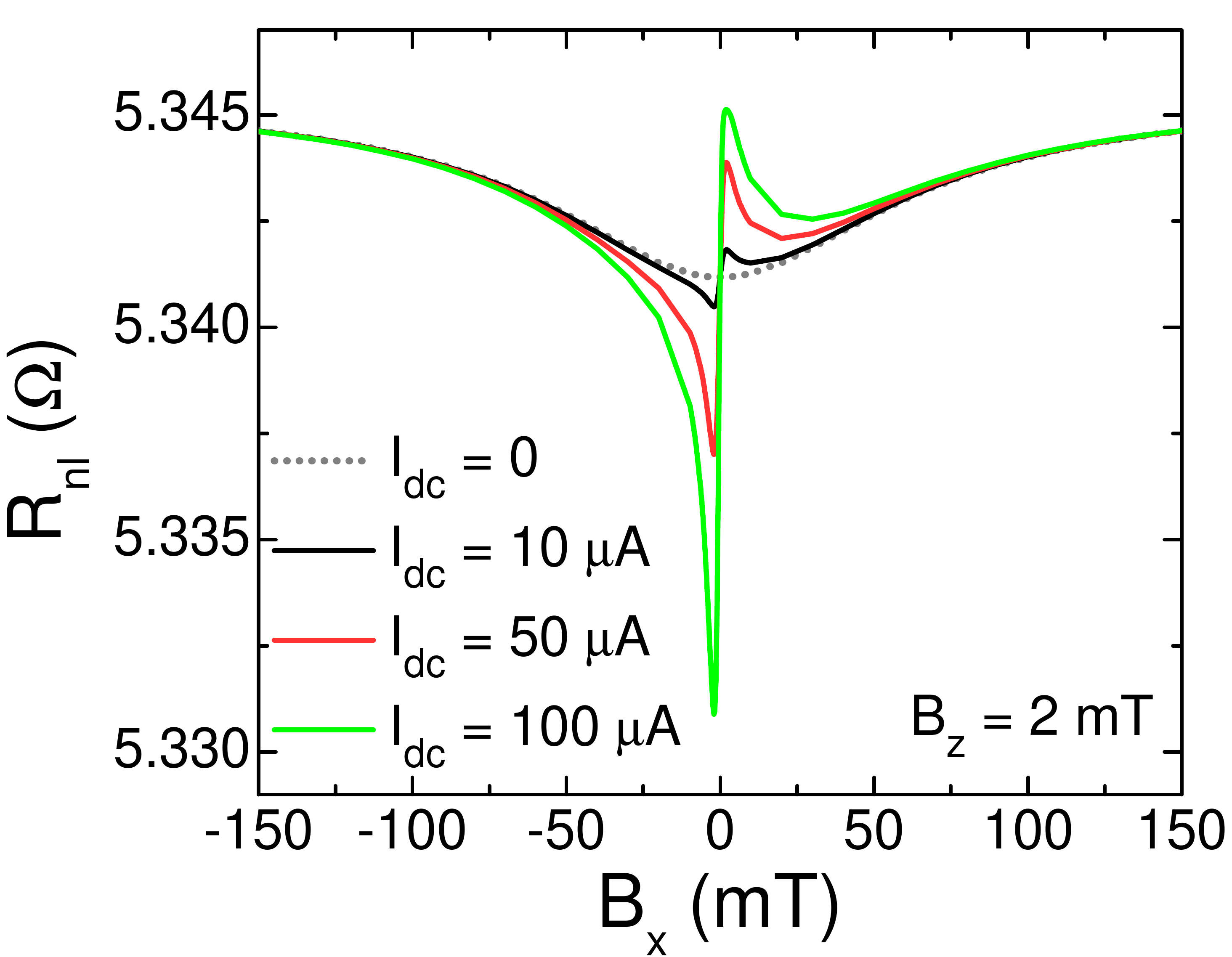} 
   \label{fig:Comsol_SV_vs_P}} 
  \subfloat[]{  
   \includegraphics[width=0.5\columnwidth]{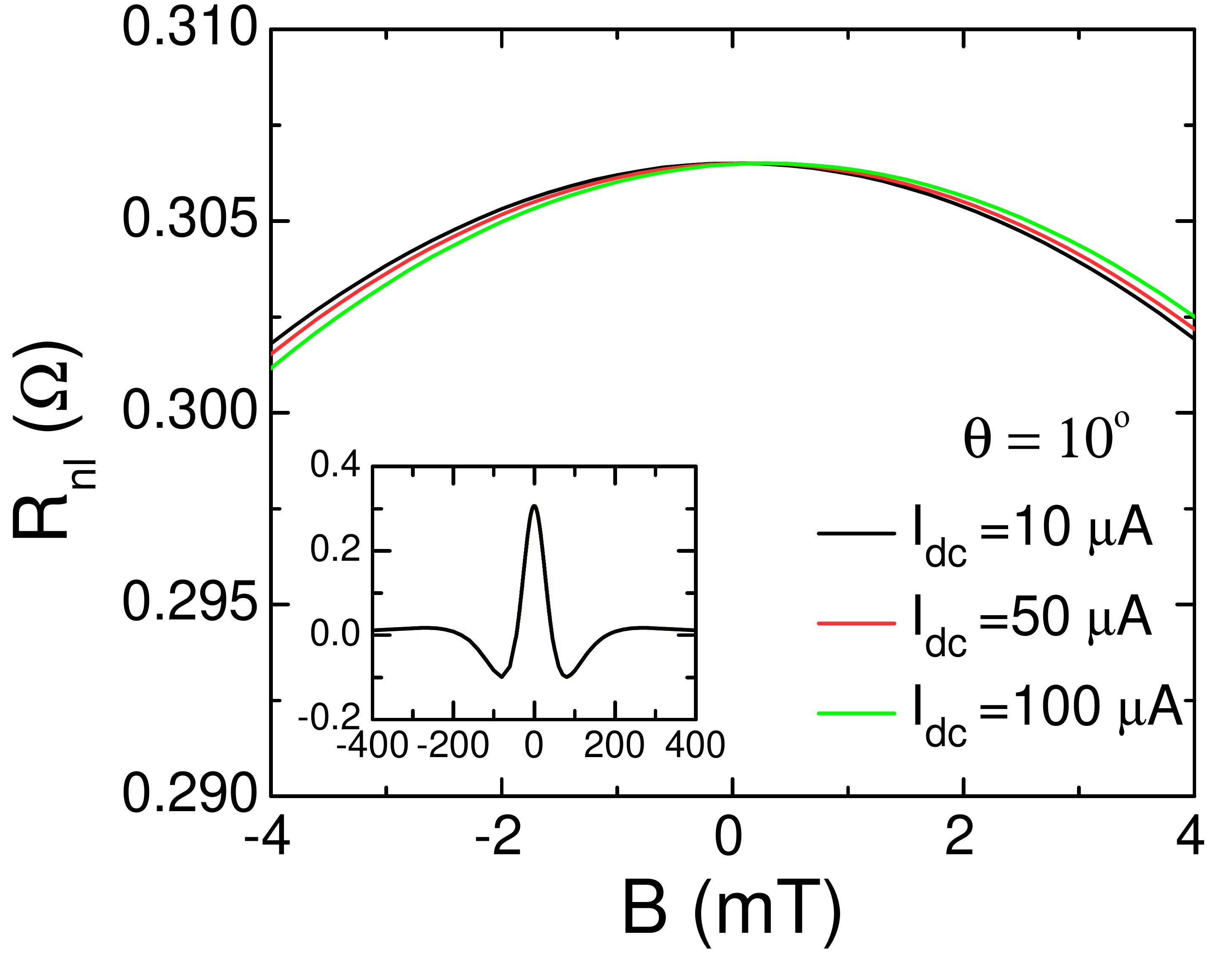} 
   \label{fig:Comsol_Hanle_vs_P}}
\captionsetup{format=hang,justification=centerlast, font=small}
\caption[]{(Color online) (a) Simulation of the non-local spin-valve signal at fixed $B_{z}$~=~2~mT under the influence of a nuclear field $\mathbf{B_{n}}$ for different injection currents $I_{\text{dc}}$ at $L$~=~1~$\mu$m in $\uparrow \uparrow$ alignment. Values of $I_{\text{dc}}=10, 50, 100$~$\mu$A correspond to the graphene polarization $\beta$=2, 10, 20\% respectively. The values of $\sigma$, $D_{s}$, $\tau_{s}$ used in the model are determined experimentally for CVD graphene. b) Simulation of the Hanle effect for magnetic field at oblique angle $\theta$~=~10$^{\circ}$ under the influence of nuclear field $\mathbf{B_{n}}$ for different polarization currents $I_{\text{dc}}$ at $L$~=~5~$\mu$m. The inset presents full Hanle curves for all $I_{\text{dc}}$, the main figure zooms to the region where the curves for each $I_{\text{dc}}$ do differ.
}
\label{fig:Comsol_SV_Hanle}
\end{figure}

Now we want to estimate the spin polarization at which the nuclear field would cause experimentally resolvable features. Typically in our CVD graphene $\lambda_{s} = 1 \mu$m, $D_{s}=0.03 $m$^{2}$/s, $\tau_{s} =100$~ps and $P=0.1$. For modeling $\mathbf{B_{n}}$ we use Eq.~(\ref{eq:Bnuclei_gr}) with $b_{n} = $-5.2~mT and we ignore any leakage effects ($f=1$), which is the best case scenario. The electronic polarization $\beta$ in graphene can be enhanced by injecting a large spin polarized dc current $I_{\text{dc}}$. $\beta$ also depends on the position of the Fermi level and therefore can be tuned by the gate voltage. Its value is largest at the Dirac point; however, it is limited by the residual carrier doping from impurities, inhomogeneities and substrate ($n_{\text{res}}\approx 10^{11}$ cm$^{-2}$), so in the simulation we take ${E_{F}} \simeq 50$~meV.
%*** CHECK number
We model the transport features in the oblique spin-valve and oblique Hanle magnetic field configurations as a function of $I_{\text{dc}}$; see Fig.~\ref{fig:Comsol_SV_Hanle}. In Fig.~\ref{fig:Comsol_SV_vs_P} we see a dip even without nuclear field $B_{n}$ (dashed line) due to a small constant out-of-plane field component $B_{z}$ (the inverted Hanle effect \cite{Dash2011}). On top of that there is a modulation due to the hyperfine effects, but even for the largest $I_{\text{dc}}=100 \mu$A this modulation is very small ($<$15~m$\Omega$) and cannot be resolved experimentally. An even smaller change due to the polarized nuclear field appears in the oblique Hanle effect [see Fig.~\ref{fig:Comsol_Hanle_vs_P}], where it leads to a minute shift of the peak position (without formation of any asymmetric peaks in the line shape; see the inset). From the simulation we can see that the nuclear effects are very small and difficult to resolve experimentally due to a very small value of $b_{n}$.

\captionsetup[subfloat]{captionskip=-1.5em,margin = 0.1em,justification=raggedright,singlelinecheck=false,font=normalsize, position=top}

\begin{figure}
\centering
    \subfloat[]{   
   \includegraphics[width=0.5\columnwidth]{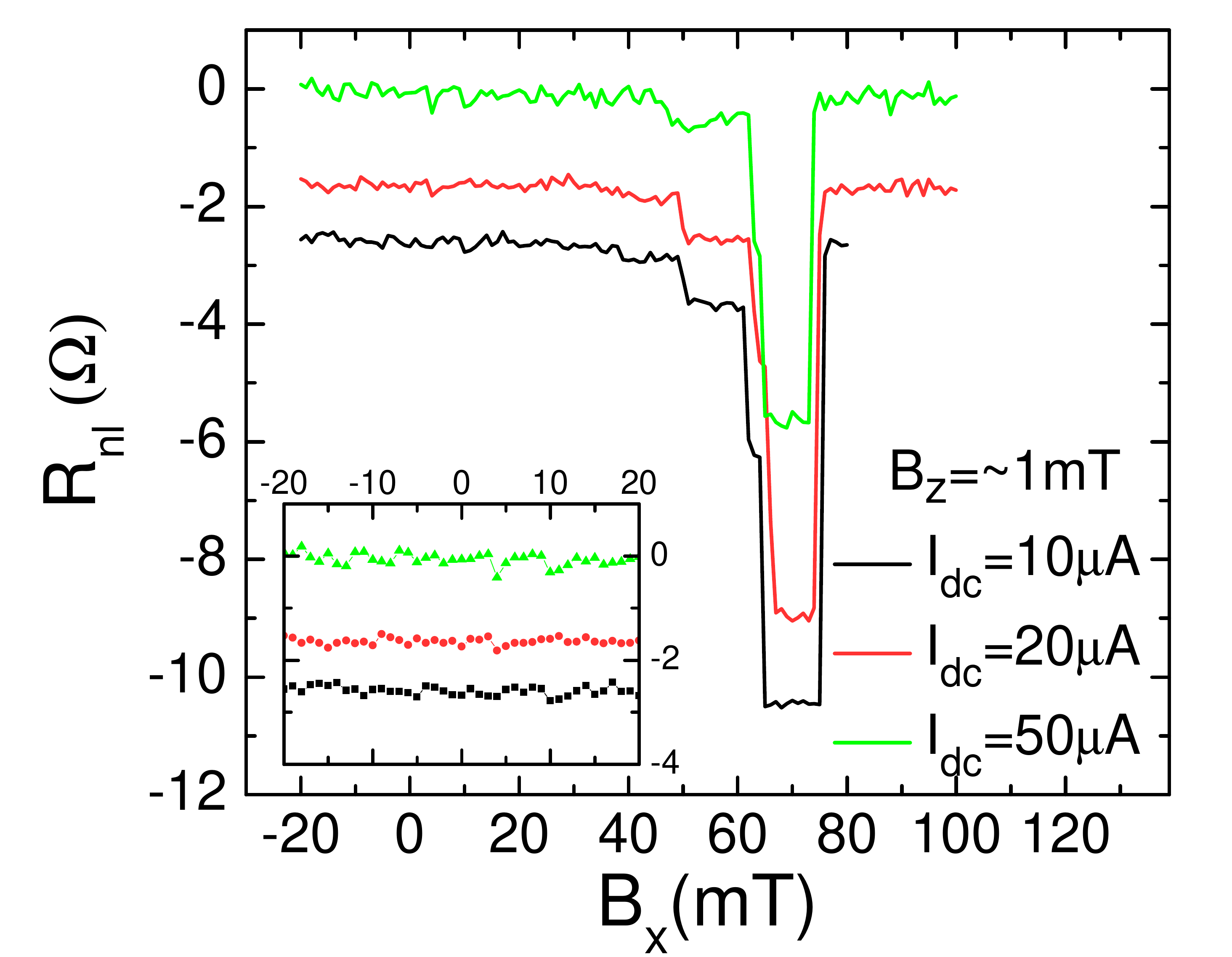} 
   \label{fig:C13_SV_vs_Idc}} 
  \subfloat[]{   
   \includegraphics[width=0.5\columnwidth]{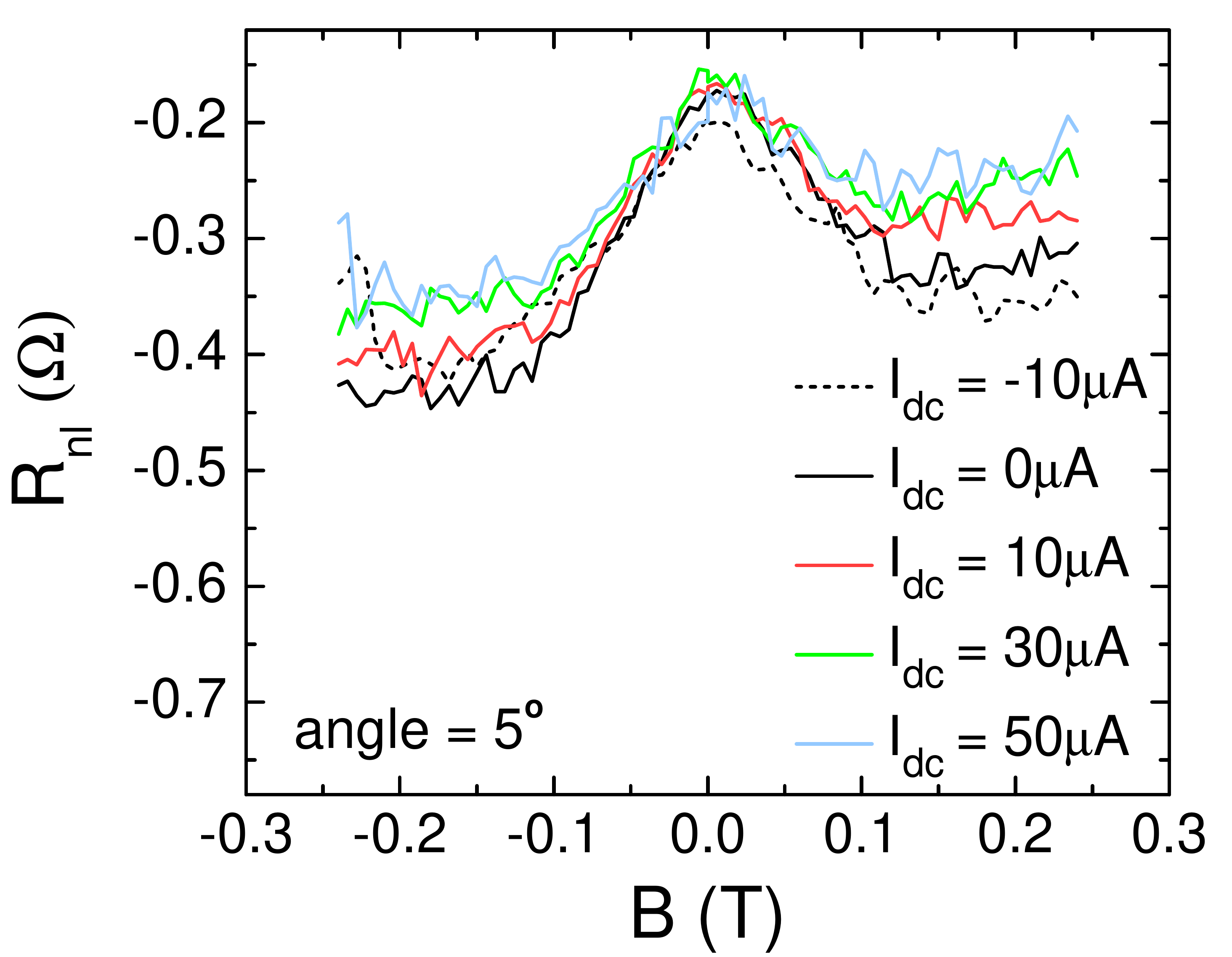} 
   \label{fig:C13_Hanle_vs_Idc}}
\captionsetup{format=hang,justification=centerlast, font=small}
\caption[]{(Color online) (a) Measurements of the non-local spin-valve signal with varying polarization currents $I_{\text{dc}}$ at fixed $B_{z}$~=~1~mT  at $L$~=~1.5~$\mu$m. The inset presents a zoom into region around $B_{x}$~=~0. No dip around zero could be observed for all strengths of polarization currents $I_{\text{dc}}$ b) Measurements of the Hanle effect for magnetic field at oblique angle $\theta$~=~5$^{\circ}$ upon varying polarization currents $I_{\text{dc}}$ at $L$~=~4~$\mu$m ($\uparrow \uparrow$). No asymmetry in Hanle lineshape could be clearly resolved. The linear background present for all $I_{\text{dc}}$ comes from the Ohmic (not spin dependent)  contribution to the non-local signal.
}
\end{figure}

To confirm experimentally the weak character of hyperfine effects we perform spin transport measurements at oblique magnetic fields and enhanced electron spin polarization. For that we send through the device a relatively large dc current (up to 50 $\mu$A) besides the small ac modulation (1 $\mu$A) used for lock-in detection.  In Fig. \ref{fig:C13_SV_vs_Idc} we show a non-local spin-valve signal under a field with small out-of-plane component $B_{z}\sim 1$~mT for varying $I_{\text{dc}}$. No dip around $B = 0$ could be observed for all strengths of polarization current $I_{\text{dc}}$ used. We should note that by sending a large dc current  on the one hand we increase graphene polarization $\beta$ but on the other hand we bias the tunneling injector and slightly decrease its spin injection efficiency (decrease of contact polarization). This can be  recognized in the decrease of the spin-valve amplitude (ac signal) for increasing $I_{\text{dc}}$, see Fig.~\ref{fig:C13_SV_vs_Idc}.  The observed decrease is relatively small  (up to 30\% of the spin signal at $I_{\text{dc}}=0$) which still maintains the increase of $\beta$ with $I_{\text{dc}}$. For the largest $I_{\text{dc}}$ the spin accumulation $\mu_{s}=e R_{\text{nl}}I_{\text{dc}}/P\simeq 3$~meV.  This results in $\beta$~=~2.5\% at the injector ($L=$~0~$\mu$m) and $\beta$~=~0.6\% at the detector $L=$~1.5~$\mu$m. 

As for conduction electrons the DNP is expected to build up very slowly ($T_{1e}\sim 10^{4}$~s), we also perform measurements of minor loops (narrower sweep of the magnetic field, where only the detector contact switches its magnetization) at constant polarizing dc current $I_{\text{dc}}=30$~$\mu$A. The measurement lasts 4~h and still no features around zero field, which could be attributed to the nuclear field, could be resolved, see Appendix \ref{sec:Minor_loop}.

\captionsetup[subfloat]{captionskip=-1.7em,margin = 0.1em,justification=raggedright,singlelinecheck=false,font=normalsize, position=top}

Next we experimentally investigate the line shape of the Hanle effect under the magnetic field at oblique angle $\theta=5^{o}$ for various $I_{\text{dc}}$ in parallel configuration, see Fig.~\ref{fig:C13_Hanle_vs_Idc}. Although in our CVD graphene $\lambda_{s}$ was relatively short, $\lambda_{s}\cong$~1~$\mu$m, thanks to high polarization of the contacts ($P=8-10\%$) it was possible to observe a spin signal even at large distances (for $L\simeq$~5~$\mu$m). No asymmetry in the Hanle line shape could be unambiguously resolved. The measured Hanle curves show only linear background from the Ohmic (not spin-dependent) contribution to the non-local signal. The scan size is limited by the magnetic anisotropy of the ferromagnetic contacts, which undergo switching at too high oblique $\mathbf{B}$.

\section{\label{sec:Discussion}Discussion}
The absence of any hyperfine-induced features in the lineshape of the spin-valve resistance or Hanle curves confirms that it is not possible to create substantial nuclear polarization by conduction electrons and that the hyperfine coupling is too weak to be measurable. The DNP in graphene, even if hyperfine interactions had a comparable strength to GaAs, could not be efficiently induced due to the lack of localized electrons (and hence small correlation times between electron and nuclei). Enhancing the electron spin polarization to increase the probability of momentum transfer from electron to nuclei also has limitations. In graphene $I_{\text{dc}}$ cannot be significantly enhanced due to the use of highly resistive tunnel contacts, which break down at large currents. 
In GaAs it is possible to achieve higher polarization [0.2-6\% (Ref. \onlinecite{Chan2009})  with $I_{\text{dc}}>$~1~mA] because of the different nature of the Schottky contacts and lower resistance of the junction. Also, graphene's thermal properties limit the maximum $I_{\text{dc}}$ (in the current annealing process one can go up to $I_{\text{dc}}\sim$~1~mA/$\mu$m before graphene breaks \cite{Tombros2011}). The upper limit for $I_{\text{dc}}$ tried here is 50~$\mu$A, which corresponds to a sizable voltage drop across the junction of  $\sim$~0.3~V and spin accumulation of $\sim$~3~meV at the contact). These differences between GaAs and graphene explain the absence of any features associated with intrinsic nuclear magnetic fields in graphene, which are very pronounced in GaAs. The presented attempt to build up and detect DNP serves as an additional experimental confirmation of the negligible size of hyperfine interactions in graphene, alongside the observation of the same $\tau_{s}$ in pure $^{12}$C- and $^{13}$C-graphene. 
  
We also attempted to induce a nuclear magnetic resonance, for which we measure the $R_{\text{nl}}$ at oblique angle and $I_{\text{dc}}$~=~50~$\mu$A, see Appendix \ref{sec:NMR}. Also this attempt to modify (here reduce due to the rf field) the spin nuclear polarization showed no effect on the spin transport, which means that the nuclear fields in graphene are negligible and/or that the dc currents used here are unable to polarize the nuclei. 

 \section{\label{sec:Conclusions}Conclusions} 
In this work we experimentally verify the role of hyperfine interactions in spin transport in graphene. We observe that the spin relaxation time in graphene is not reduced by hyperfine interactions; even when we compare fully isotopic $^{13}$C- against fully isotopic $^{12}$C-graphene.  
Further, we perform a set of experiments in various configurations to amplify the hyperfine effects. In oblique spin-valve and Hanle measurements we tried to observe features of dynamically induced nuclear polarization, by creating a sizable electron polarization in graphene of $\sim$2.5\%, but no distinctive features related to nuclei are observed. With the finite-element method we model the spin Bloch equation for graphene at oblique angles and we are able to estimate the lower limit for graphene polarization to result in any measurable fingerprints of the nuclear magnetic field. Even for the highest achievable spin polarization in graphene, the hyperfine features cannot be experimentally resolved. This is further confirmed by the measurements at oblique angle at high polarizing currents. 
This paper experimentally proves the negligible role of the intrinsic hyperfine interactions in graphene for spin relaxation, in agreement with theory. Yet the possibility of observing a spin signal over relatively large distances in CVD graphene   confirms the choice of graphene as an efficient spin transport material for future applications.

\medskip
% %
\begin{acknowledgments} 
We would like to thank P.~Crowell, G.~Salis, G.~Chad and A.~R.~Onur for useful discussions. We would like to acknowledge M.~H.~D.~Guimaraes,  H.~M.~de~Roosz, J.~Holstein and B.~H.~J.~Wolfs for technical support. This work was financed by NanoNed, the Zernike Institute for Advanced Materials and the Foundation for Fundamental Research on Matter (FOM). %\end{acknowledgments}
\end{acknowledgments}

\onecolumngrid
\appendix

\section{\label{sec:Spin_valve_polarization_changes} Comparison of spin properties of $^{13}$C-graphene at room temperature and T~=~4.2~K.}

\begin{figure}[!h] 
\centering 
\includegraphics[width=0.7\columnwidth]{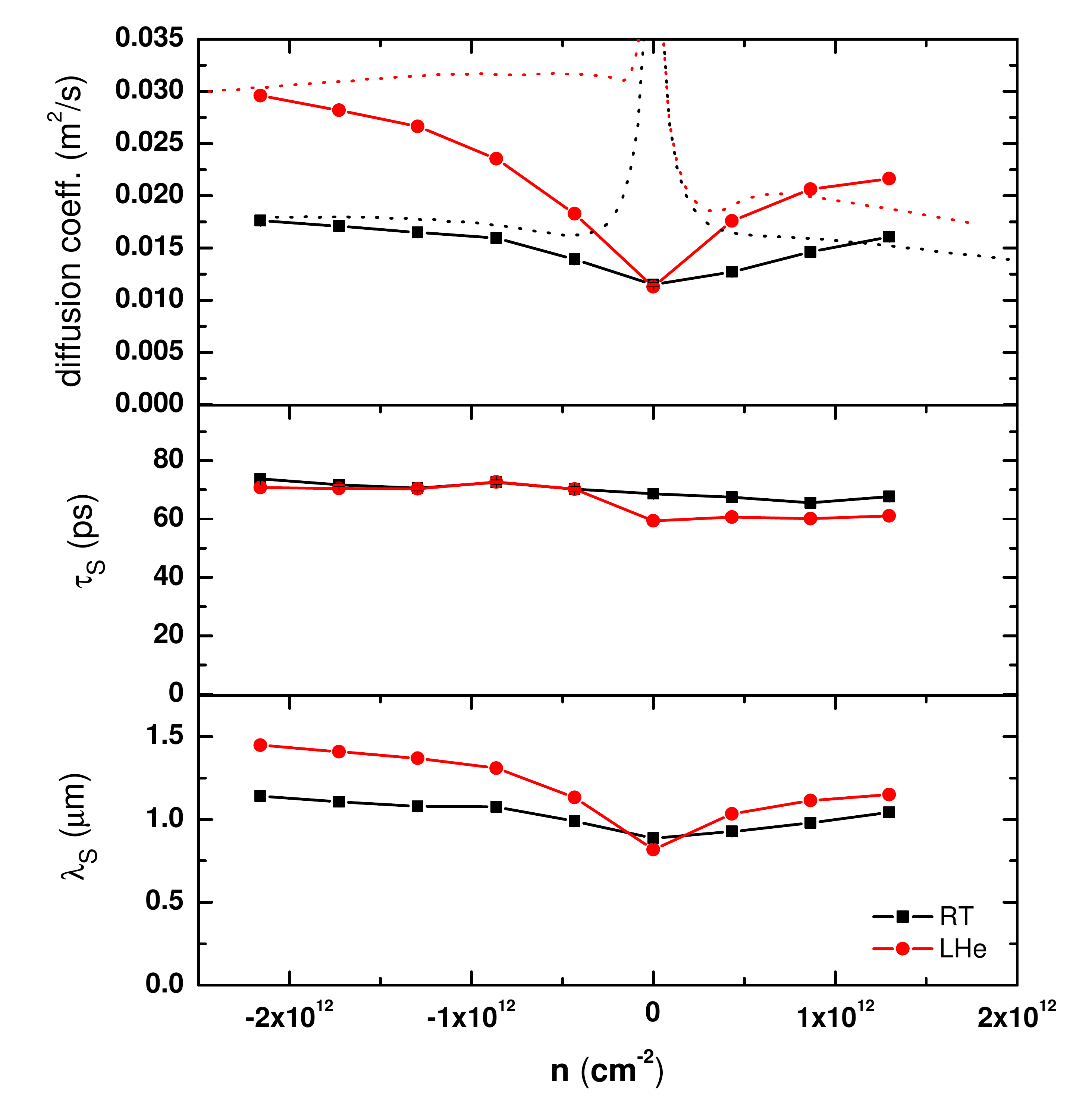} \captionsetup{format=hang,justification=centerlast, font=small} 
\caption[]{(Color online) Comparison of transport properties  at room temperature and at T~=~4.2~K as a function of carrier concentration. Top panel: diffusion coefficient, central panel: spin relaxation time $\tau_{s}$, bottom panel: spin relaxation length $\lambda_{s}$. Dotted line represents $D_{c}$, calculated using the Einstein relation (from charge transport), $D_{s}$, $\tau_{s}$ are extracted from the Hanle curves with subtracted background and $\lambda_{s} = \sqrt{D_s \tau_s}$. } 
\label{fig:C13_Hanle_RTvsLHe} 
\end{figure}

Hyperfine effects are most pronounced at low temperatures due to less thermal fluctuations of nuclear spin, hence we want to compare the spin transport properties at room temperature versus T~=~4.2~K, extracted from Hanle measurements for parallel and antiparallel contacts configuration as a function of gate voltage. The fitting of Hanle precession curves gives independent values for spin relaxation time $\tau_{s}$ and spin diffusion $D_{s}$, from which we get $\lambda_{s}=\sqrt{\tau_{s} D_{s}}$. All these coefficients are summarized in Fig.~\ref{fig:C13_Hanle_RTvsLHe}. Alternatively, from the charge transport measurements we can determine charge diffusion constant $D_{c}$ using the Einstein relation $\sigma = e \nu(E)D_{c}$ , where $\nu(E)$ is the density of states of graphene at $T$~=~0~K and $\sigma$~=~1/$\rho$ is its sheet conductivity.  The singularity of $D_{c}$ around $n$~=~0 arises due to the vanishing number of states at the Dirac point and can be eliminated by including the broadening of states \cite{Jozsa2009}. Nevertheless, as these corrections are negligible at the large doping, in metallic regime one can rely on $D_{c}$ determined using the ideal, zero-temperature density of states. We can see that for large $n$ $D_{c}\simeq D_{s}$, like in exfoliated graphene \cite{Jozsa2009}. 
In graphene with intrinsic magnetic fields from defects \cite{McCreary2012, Birkner2013}, there appears an extra scaling of the precession term $\frac{g \mu_B}{ \hslash} \boldsymbol{B}$, because the external magnetic field, used in the fitting procedure, is different from the total magnetic field experienced by spins. 
Such a scaling can be also seen as a change in the graphene $g$-factor, $g\rightarrow g^{*}$, and it affects the determination of spin coefficients: $\tau_{s}=\frac{g^{*}}{g}\tau_{s}^{*}$, $D_{s}=\frac{g}{g^{*}}D_{s}^{*}$. This similarity between $D_{s}$ and $D_{c}$, determined by two independent methods, implies that there is no change in $g$-factor due to internal (here nuclear) fields, and the spin coefficients are properly determined. The values for $\tau_{s}$ range from 60 to 100~ps, depending on the doping, and they barely change with the temperature. Similar weak dependence of $\tau_{s}$ on temperature is also present in exfoliated graphene, although $\tau_{s}$, determined from Hanle precession, is 2-10 times higher \cite{Tombros2007, Han2011}. 
The observed lower values of $\tau_{s}$ are also found in the control sample - a CVD graphene from pure $^{12}$C precursor and therefore cannot be attributed to the hyperfine effects. A lower $\tau_{s}$ can originate from crystal defects and rippling of graphene sheet, which is inherent to the growth and transfer conditions, the quality of the Cu substrate and eventual remaining of FeCl$_{3}$ etchant.

\section{\label{sec:Modeling_GaAs} Simulation of oblique effects in GaAs.}
In standard Hanle precession experiment the magnetic field is set perpendicular to graphene plane $\boldsymbol{B}=(0,0,B)$, thus the precession term $\boldsymbol{B}\times \boldsymbol{\mu_{s}}$ of Bloch equation (Eq.~1 in the main text) vanishes in $z$-direction. Additionally, the magnetization of injecting contact, and in consequence orientation of injected spins, lies in the $x$-$y$ plane. Therefore, one can simplify the description of spin dynamics by considering the spin accumulation vector only in two-dimensions: $\boldsymbol{\mu_{s}} = (\mu_{s,x}, \mu_{s,y})$. However, for a general orientation of external magnetic field $\boldsymbol{\mu_{s}}$ needs to be considered in all 3 directions. To observe the electron spin dephasing from the nuclear field it is necessary to apply an external field at an oblique angle, which will redirect the nuclear magnetization from its collinear alignment to the electron spin. Additionally, the mutual dependence of hyperfine field $\boldsymbol{B_{n}}$ and spin accumulation  $\boldsymbol{\mu_{s}}$ (Eq.~3 in the main text) leads to a non-linear term in Bloch equation, and requires self-consistent solving. 
These two aspects: a need for a general 3D form of Bloch equation and non-linear dependence of $\mu_s$  make the predictions of the spin signal very difficult. To tackle this problem we use a finite-element software package (COMSOL Multiphysics, version 4.3) which allows to define a set of partial differential equations (PDE) and solve it for an user defined geometry. We use a two channel model \cite{Valet1993} for spin transport, in which we define spin-dependent chemical potential as a variable $(\boldsymbol{\mu_{\uparrow}}, \boldsymbol{\mu_{\downarrow}})$, where $\boldsymbol{\mu_s}=\boldsymbol{\mu_{\uparrow}}-\boldsymbol{\mu_{\downarrow}}$.  
In arbitrary external magnetic fields, not necessary perpendicular to the graphene plane, the electron spins change their orientation, therefore we define the chemical potential in 3D spin space separately for each spin channels: $\boldsymbol{\mu_{\uparrow\!/\downarrow}} = (\mu_{x}^{\uparrow/\downarrow}, \mu_{y}^{\uparrow/\downarrow}, \mu_{z}^{\uparrow/\downarrow})$. The up and down arrows refer to the spin orientation only at the point of injection and not at the distances further away from injector, where spin orientation undergoes precession. We emulate the spin dynamics separately for each spin channel, linking them only by spin relaxation.
In this problem the charge and spin transport are coupled, therefore when solving the diffusion equation  the conservation of generalized currents $\triangledown (\boldsymbol{J_{\uparrow}},\boldsymbol{J_{\downarrow}}) = f(\boldsymbol{\mu_{\uparrow}}, \boldsymbol{\mu_{\downarrow}})$ has to be satisfied. The conservation of charge current is given by $\triangledown (\sigma_{\uparrow} \triangledown  \boldsymbol{\mu_{\uparrow}}+\sigma_{\uparrow} \triangledown  \boldsymbol{\mu_{\downarrow}}) = 0 $. In non-magnet $\sigma_{\uparrow}=\sigma_{\downarrow}$, while in ferromagnet $\sigma_{\uparrow}=(1+P)(\sigma_{\uparrow}+\sigma_{\downarrow})$, $\sigma_{\downarrow}=(1-P)(\sigma_{\uparrow}+\sigma_{\downarrow})$, where $P$ is a polarization of a ferromagnetic contact. The conservation of spin is  derived from Valet-Fert equation with additional precession term $\triangledown^{2}\boldsymbol{\mu_s} = \boldsymbol{\mu_s}/\lambda^{2}-\boldsymbol{\omega}\times \boldsymbol{\mu_s}/D_s$, where the term $\boldsymbol{\omega} =\frac{g \mu_B}{ \hslash} (\boldsymbol{B}+\boldsymbol{B_{n}})$ includes external and nuclear magnetic field. The 3D COMSOL model includes full geometry of the device together with ferromagnetic contacts and the tunnel barriers of finite resistance. 

To verify our model we choose GaAs system and spin transport results of Chan \emph{et al.}\cite{Chan2009} as a reference. We set the device geometry corresponding to the one reported (meshed with tetrahedrons) and set charge and spin transport coefficients from the values determined therein: $b_n$~=~-5.3~T, $b_e$~=~-5~mT, $\tau_s$~=~10~ns, $\lambda_s$~=~5~$\mu$m, $L=10$~$\mu$m,  GaAs polarization $\beta_{\text{GaAs}}=\frac{3\mu_{s}}{E_{F}}$, where $E_F\simeq 7.5$~meV, $P=$~0.2. 
\captionsetup[subfloat]{captionskip=-0.4em,margin = 0.1em,justification=raggedright,singlelinecheck=false,font=normalsize, position=top}

\begin{figure}[!htb]
\centering
    \subfloat[]{   
   \includegraphics[width=0.45\columnwidth]{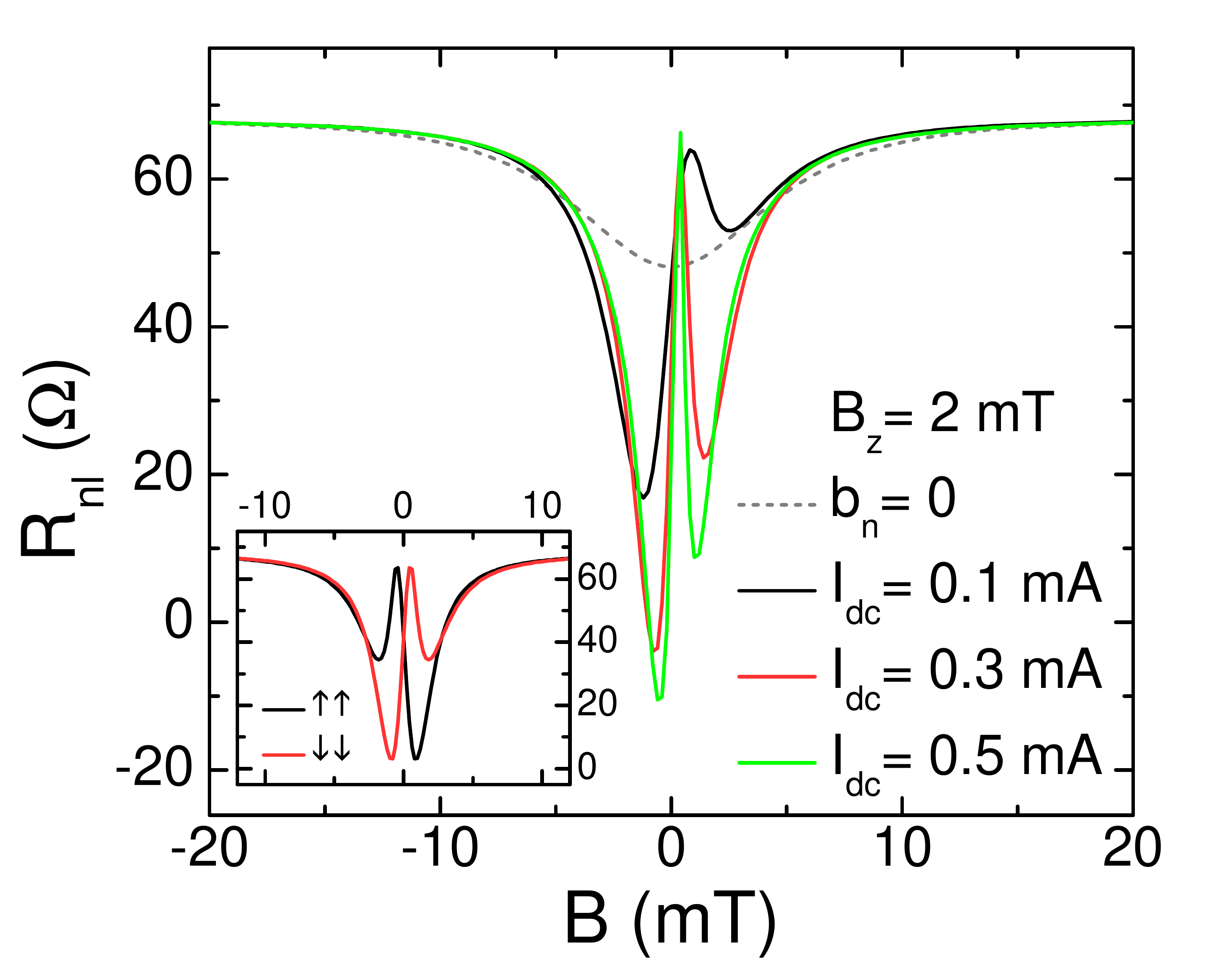} 
   \label{fig:GaAs_SV_vs_Idc}} 
  \subfloat[]{   
   \includegraphics[width=0.45\columnwidth]{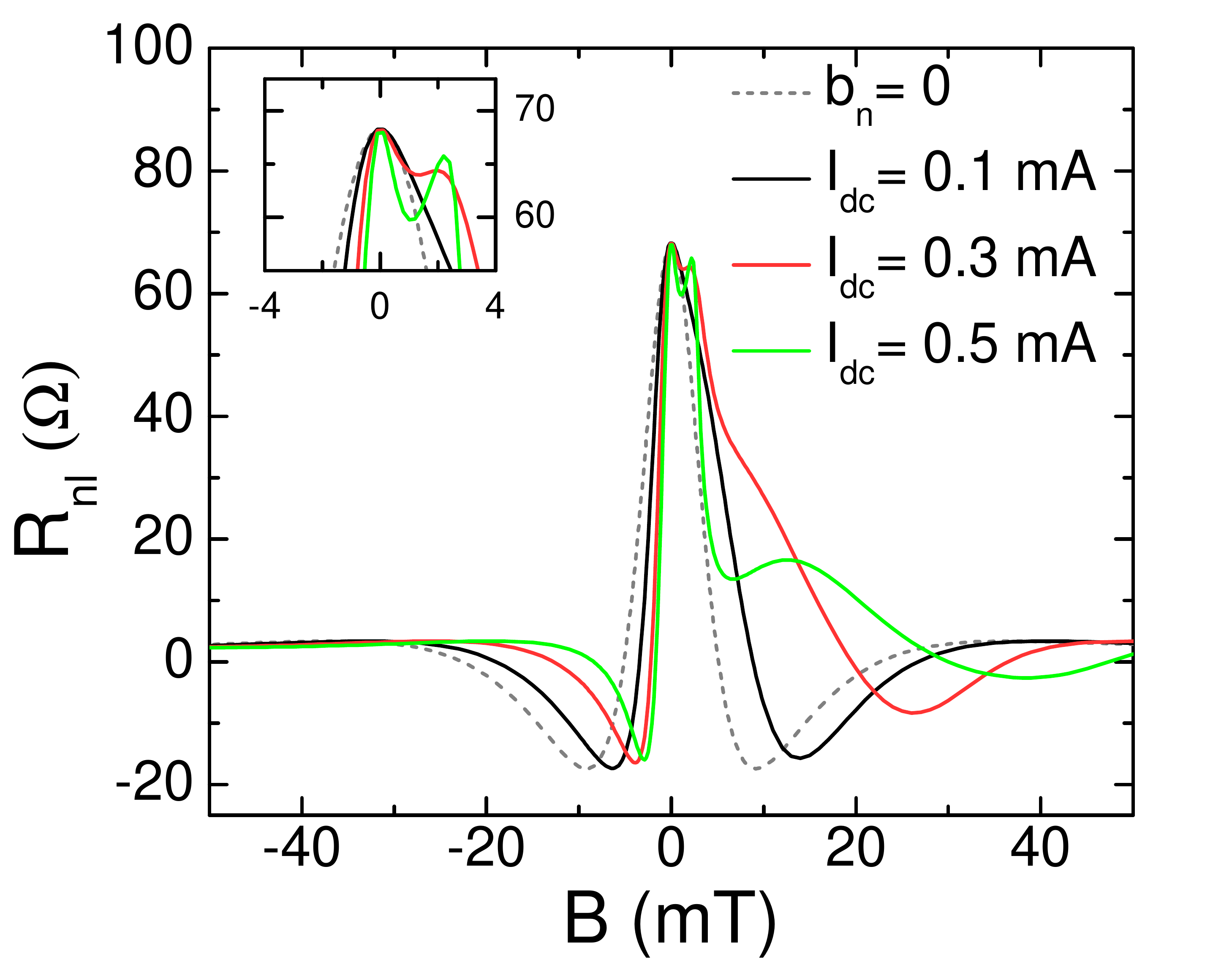} 
   \label{fig:GaAs_Hanle_vs_Idc}}
\captionsetup{format=hang,justification=centerlast, font=small}
\caption[]{(Color online) (a) Simulated oblique non-local spin signal in presence of hyperfine field in GaAs for different values of polarizing current $I_{\text{dc}}$. The increase of polarization translates to increase of depolarization dips around $B=0$. The dashed line presents the signal in case of no hyperfine interactions ($b_{n}=0$). The inset shows mirroring property of the spin signal in parallel contact configuration between different magnetization directions ($\uparrow \uparrow$ or $\downarrow \downarrow$). b) Simulated oblique Hanle precession signal for $\theta=10^{o}$. With the increase of polarization $I_{\text{dc}}$ the central Hanle peak shifts it position and an asymmetry builds up. The inset zooms at the $B=0$, where the central peak splits. }
\end{figure}

We first emulate the lineshape of the GaAs spin valve signal in presence of small, fixed, out-of-plane field component $B_z$~=~2~mT as a function of polarizing current $I_{\text{dc}}$, see Fig.~\ref{fig:GaAs_SV_vs_Idc}. For parallel contact magnetization  ($\uparrow \uparrow$) we observe a formation of the depolarization dip around $B_{x}$=0, of which amplitude increases with electron spin accumulation. A plot of the expected signal in the absence of nuclear polarization ($b_{n}=0$, dashed line) shows a broad, single dip, corresponding to the inverted Hanle effect due to the presence of small constant out-of-plane field component $B_{z}$. The structure of the hyperfine dips mirrors along $B=0$ axis when the polarization of contacts reverses (from $\uparrow \uparrow$ to $\downarrow \downarrow$), see inset in Fig.~\ref{fig:GaAs_SV_vs_Idc}. This is directly related to the change in the vectorial configuration of the magnetic fields involved and was also confirmed experimentally \cite{Chan2009}.

Next we simulate the oblique Hanle effect by setting an external field at oblique angle $\theta$~=~10$^{o}$, so $\boldsymbol{B}=(B\sin{\theta}, 0, B\cos{\theta})$. 
With an increase of $I_{\text{dc}}$ a central Hanle peak shifts to the positive field values, creating an asymmetric lineshape, see Fig~\ref{fig:GaAs_Hanle_vs_Idc}. Additionally, the maximum peak at $B$=0 splits into two, see inset in  Fig~\ref{fig:GaAs_Hanle_vs_Idc}. This complex lineshape is a result of subtle interplay between all the magnetic fields included in the problem. These features are less pronounced than the dips in the spin-valve signal and require much higher spin accumulation to be resolved.

All the obtained features are in agreement with the experimental findings of Chan \emph{et al.}\cite{Chan2009}, confirming the validity of our model.

\section{\label{sec:Minor_loop} Building up nuclear polarization - minor loop scan.}
\begin{figure}[!htb] 
\centering 
\includegraphics[width=0.7\columnwidth]{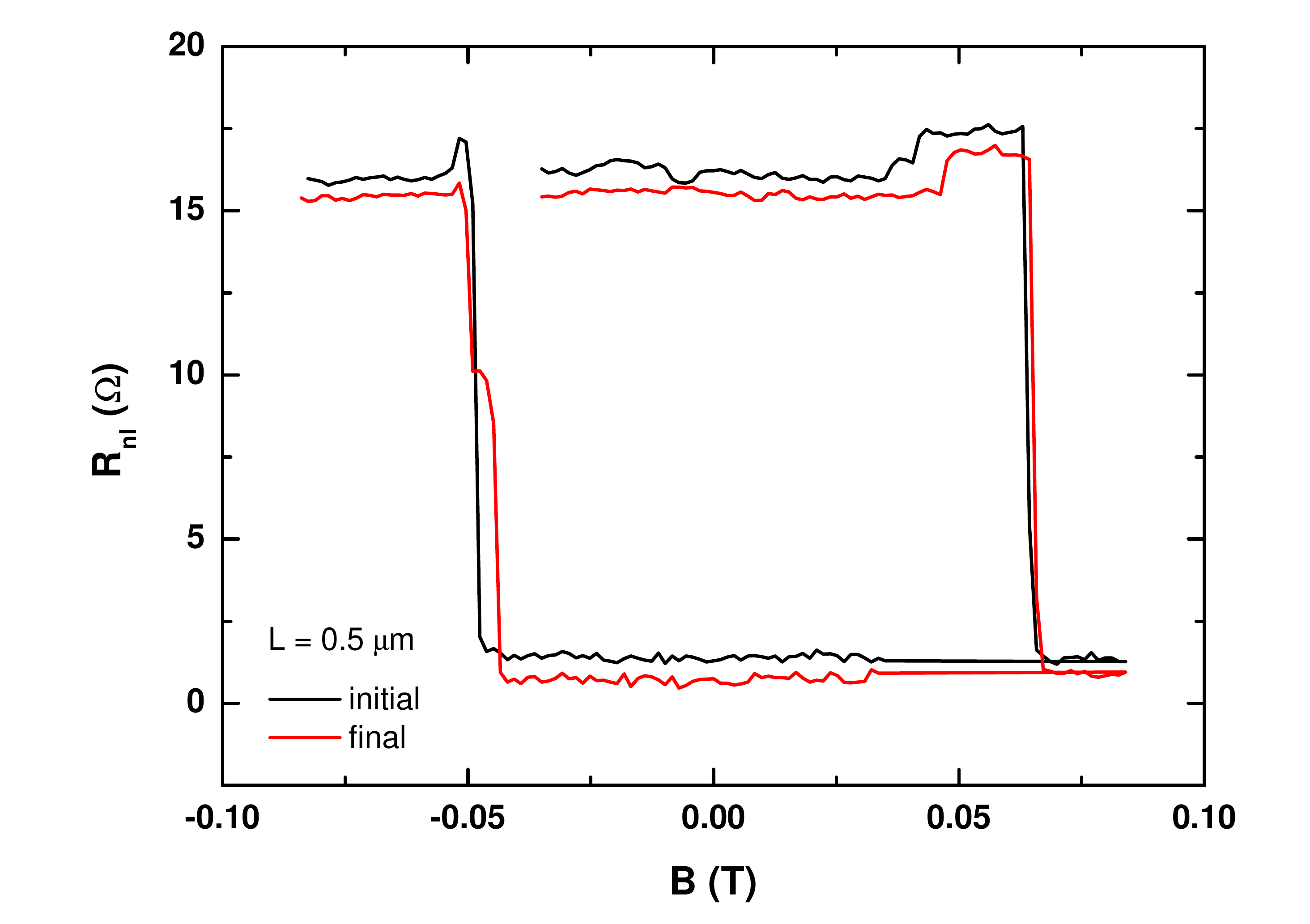} \captionsetup{format=hang,justification=centerlast, font=small} 
\caption[]{(Color online) First and last (50$^{th}$) measurement of minor loop from a set of 50 consecutive scans. The distance between injector and detector L~=~0.5~$\mu$m, the polarizing current $I_{\text{dc}}$~=~30~$\mu$A, at T~=~4.2~K.   } 
\label{fig:C13_Minor_loop} 
\end{figure}

As motivated in the main text the correlation time between delocalized conducting electrons and nuclei is very short and requires times of the order $\sim 10^4$~s to build up dynamical nuclear polarization. For that reason we perform a spin-valve measurement in a reduced magnetic field range, such that only the detector contact switches its magnetization (so called minor loop measurement). This way the injected polarizing current $I_{\text{dc}}=30 \mu$A has always the same spin direction. The eventual formation of dephasing dips in $R_{\text{nl}}$ around zero field is monitored as a function of time by performing continuous field sweeps. We measure the effect at the closest distance between injector-detector, where the spin accumulation is the highest. We perform 50 sweeps within the total time of 250 minutes ($\simeq 1.5 \cdot 10^{4}$~s), the first and the last sweep can be seen in Fig.~\ref{fig:C13_Minor_loop}. The scans are identical except for the shift in common background. No extra dips around B=0, which could be attributed to polarized nuclei can be resolved.

This, as well as other measurements reported here, confirms the negligible role of hyperfine interactions between electrons and nuclei in graphene.

\section{\label{sec:NMR} NMR studies of the graphene transport.}
As a last check of the possible correlation between nuclear field and spin dephasing mechanisms we try to induce nuclear magnetic resonance in $^{13}$C nuclei. Upon the high frequency electromagnetic field, which energy matches the energy difference between the nuclear spin levels, we can induce transitions between these states, thus enhancing the randomization of the nuclear field. This causes the change in the effective magnetic field acting on electron spins from polarized nuclei, which could affect the spin signal provided that the nuclear polarization and hyperfine interactions are sufficiently strong. The resonance frequency for $^{13}$C nuclei appears at 10.7~MHz/T \quad \cite{Harris1978}. 
% In thermal equilibrium in magnetic field nuclear spins I=1/2 populate two Zeeman levels p+ and p-, where p+ + p- =1. The ratio p+/p- = exp(gamma*hbar*H/kBT), which in the high temperature approximation gives: p+-p- = gamma*hbar*H/2kBT. 
% Bcause the probablitily of a transition induceed by the r.f. field is equal to that of opposite transition, the r.f. field tends to decrease the difference between the populations below its thermal equilibrium value. P'+ and P'- are result of competition between the r.f. irradiation and spin-lattice relaxation. We can define a 'spin' temperature T_S higher than the lattice temperature, by analogues relation P'+/P'- = exp(gamma*hbar*H/kBT_S). In particular an infinite spin 'temperature' T_S will describe a situation where P'+ and P'- have been made equal. This picture is not rigorously correct as we describe vectorial spin with scalar quantity - temperature and we ommit the perpendicular field components from r.f. on angular momentum. 
\begin{figure}[!htb] 
\centering 
\includegraphics[width=0.7\columnwidth]{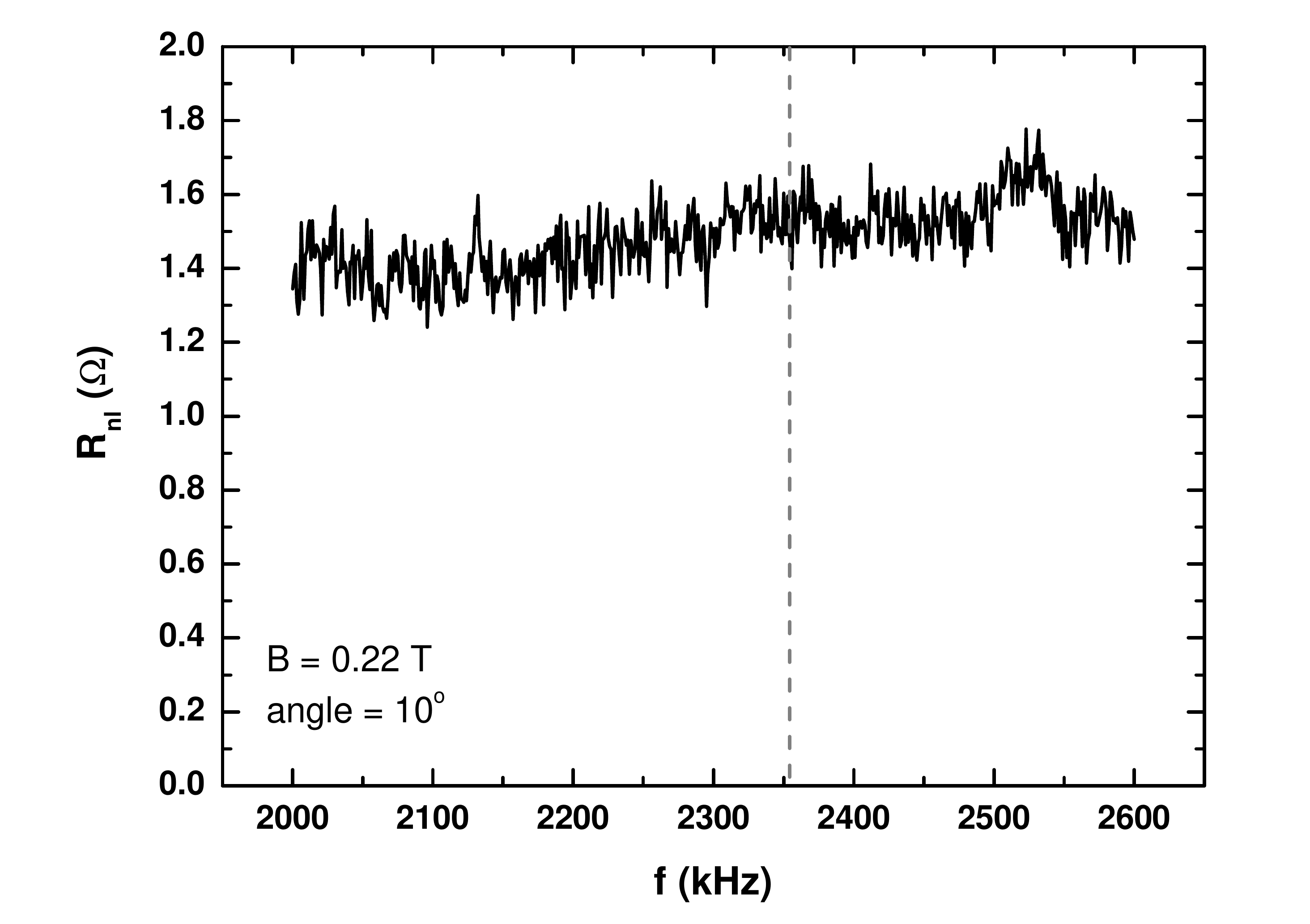} \captionsetup{format=hang,justification=centerlast, font=small} 
\caption[]{(Color online) A non-local resistance as a function of applied rf frequency for parallel contacts magnetization at T~=~4.2~K and $\boldsymbol{B}$=~0.22~T at oblique angle $\theta$=~10$^{\circ}$.  In this configuration we expect a resonance (and increase of $R_{\text{nl}}$) at $\nu_{^{13}C}$~=~2.355~MHz, indicated by the dashed line, however, no resonance peak could be distinguished. The frequency step is 1~kHz. } 
\label{fig: NMR_C13g_L=1um} 
\end{figure}

To investigate this effect we fabricated a device with a waveguide in its close vicinity to exert an rf modulation of nuclear magnetic moment. Then, at $T$~=~4.2~K, we measure the non-local spin-valve signal $V_{\text{nl}}^{\uparrow\uparrow}$ as a function of rf frequency $f$, when crossing the resonance. We set $\boldsymbol{B}$=~0.22~T, which corresponds to a resonance frequency $\nu_{^{13}C}$~=~2.355~MHz. We set the field at the oblique angle $\theta$=10$^\circ$ and inject a large polarizing current $I_{\text{ac}}$~=~50~$\mu$A together with a small ac component ($I_{\text{ac}}$~=~1~$\mu$A) for lock-in detection ($L$~=~1$\mu$m). This way we enhance the possible nuclear polarization so that its randomization under resonance conditions could be detected. If the polarized nuclei induce spins dephasing, then upon NMR and the decrease of nuclear polarization the electron spin signal should increase. We apply  the rf field of 5~dBm power and vary it with a step of 1~kHz (the rf field $B_{\text{rf}}\lesssim$1~mT). The obtained curve, see Fig.~\ref{fig: NMR_C13g_L=1um}, shows no apparent features when sweeping across the resonance frequency, indicated by dashed line. This supports again the observation of negligible influence of nuclear magnetic moment for spin dephasing in graphene.

\end{document}